# Autonomous synthesis of thin film materials with pulsed laser deposition enabled by in situ spectroscopy and automation

*Sumner B. Harris\*, Arpan Biswas, Seok Joon Yun, Christopher M. Rouleau, Alexander A. Puretzky, Rama K. Vasudevan, David B. Geohegan, and Kai Xiao\**

Center for Nanophase Materials Sciences, Oak Ridge National Laboratory, Oak Ridge, Tennessee 37831, United States.

\*Correspondence should be addressed to: harrissb@ornl.gov or xiaok@ornl.gov

**Abstract**

Synthesis of thin films has traditionally relied upon slow, sequential processes carried out with substantial human intervention, frequently utilizing a mix of experience and serendipity to optimize material structure and properties. With recent advances in autonomous systems which combine synthesis, characterization, and decision making with artificial intelligence (AI), large parameter spaces can be explored autonomously at rates beyond what is possible by human experimentalists, greatly accelerating discovery, optimization, and understanding in materials synthesis which directly address the grand challenges in synthesis science. Here, we demonstrate autonomous synthesis of a contemporary 2D material by combining the highly versatile pulsed laser deposition (PLD) technique with automation and machine learning (ML). We incorporated *in situ* and real-time spectroscopy, a high-throughput methodology, and cloud connectivity to enable autonomous synthesis workflows with PLD. Ultrathin $WSe_2$ films were grown using co-ablation of two targets and showed a 10x increase in throughput over traditional PLD workflows. Gaussian process regression and Bayesian optimization were used with *in situ* Raman spectroscopy to autonomously discover two distinct growth windows and the process-property relationship after sampling only 0.25% of a large 4D parameter space. Any material that can be grown with PLD could be autonomously synthesized with our platform and workflows, enabling accelerated discovery and optimization of a vast number of materials.

**Introduction**

Modern science has entered an era of accelerated discovery, caused by advances in automated experiments, AI, and high-performance computing (HPC)[1]. Leveraging the successes of these technologies in synthesis science is crucial to accelerate materials discovery, achieve mechanistic control over synthesis, and integrate theoretical and *in situ* characterization tools to direct synthesis in real-time[2]. Autonomous synthesis platforms have great promise to drive these critical research areas by rapidly and efficiently exploring parameter spaces to achieve accelerated understanding of process-property relationships or to direct synthesis using multimodal real-time diagnostic data. So far, significant advances in solution-based methods are realized in numerous studies[3–6] by integrating AI with automated benchtop flow reactors[7,8] or robotic laboratories[9,10] to autonomously explore large parameter spaces with minimal human intervention. In all cases, autonomous synthesis with these systems is enabled by their available suite of characterization tools like optical absorption, photoluminescence (PL), nuclear magnetic resonance (NMR), or mass spectroscopy from which material properties and optimization targets can be derived. Solution-based synthesis methods are also particularly well-suited to closed loop, high-throughput experimentation using commercial tools that have been available/developing for decades[11], particularly in the pharmaceutical industry[12]. In contrast, there are very few examples to date of autonomous synthesis using chemical or physical vapor deposition techniques (CVD or PVD), largely due to the relative lack of commercially available tools that integrate multiple diagnostics with these growth techniques and the expense associated with developing such tools from scratch.

Leading in the CVD approach is the Autonomous Research System (ARES) that demonstrated the autonomous growth of carbon nanotubes *via* laser heating of catalyst films on specialized Si micro-pillar substrates that allowed numerous experiments on a single chip and showed a 100x increase in throughput over traditional CVD[13–15]. While producing microscale samples in specialized growth apparatus' is highly enabling for rapid exploration, "lab scale" samples (e.g. 5x5 mm substrate) are preferred to facilitate many standard characterization techniques like XRD, XPS, or electrical and magnetic property measurements into autonomous workflows. PVD techniques produce lab scale samples but suffer from workflow bottlenecks related to vacuum chambers, substrate preparation, and sample transfer to analytical instruments, making true autonomous synthesis difficult to achieve. A recent example of autonomous PVD addresses these challenges with a cluster-system approach in which the deposition chamber was linked to characterization tools and used a robotic arm to transfer samples *in vacuo*[16]. The cluster-system model for PVD is promising because it is relatively common to link vacuum techniques in this way, i.e., combined molecular beam epitaxy (MBE) or PLD with XPS, STM, LEED. However, such systems are highly complex and difficult to automate[17] and existing legacy clusters may be impossible to update without a significant infrastructure investment. So far, there are no working examples of autonomous PLD or MBE systems, likely due to these automation challenges.

Common to both autonomous solution-based and CVD/PVD systems is the generation of large amounts of multi-modal data and, in an active learning setting, the need to repeatedly train AI models. Networking autonomous labs to cloud-based storage and HPC resources enables the building of openly available synthesis-characterization databases and the ability to train expensive AI models, incorporate *ab initio* calculations in autonomous workflows[18], and enable remote control over synthesis processes. Such network ecosystems have been adopted by light sources[19] and advanced microscopy labs[20,21] but the human-driven and data-sparse PVD lab environments have had little incentive to do the same, further hindering advancement in this field.

In the present work, we address priority research needs in synthesis science and transformative manufacturing through the development of autonomous synthesis at the lab scale with PLD, which is a highly versatile materials exploration and discovery technique. Our previous work already showed how *in situ* diagnostics like plume imaging can be used to measure and control key plasma parameters like kinetic energy (KE) to reveal growth windows with ex situ characterization[22] and *in situ*/real-time characterization[23]. With the digital precision and kinetic energy control of PLD, it can explore the evolution and perfected synthesis of metastable states in PLD, such as Janus structures and fractional Janus alloys[23]. Therefore, we develop an "Auto-PLD" platform equipped with a variety of both gas-phase and substrate *in situ* diagnostics into a fully automated experimental system that integrates real-time diagnostic data acquisition and monitoring with high-throughput methodology, data analysis, and machine learning in a cloud-based network ecosystem to enable autonomous materials exploration and workflow development.

As a model system for autonomous synthesis, we grow ultrathin $WSe_2$ films by co-ablation of two PLD targets, a $WSe_2$ target and a Se target (which is added to address the chalcogen loss that tend to occur in PVD). In PLD, changing the background gas pressure is often performed to adjust the KE of the species reaching the substrate, however this changes the deposition rate and it can change the nature of the species reaching the substrate from atomic and molecular species to clusters and nanoparticles[24]. Thus, while PLD is conceptually simple, the synthesis and processing environment is actually quite complex and dynamic, especially with two interacting plasma plumes. Thus, the nonequilibrium processing advantages of PLD (digital delivery, tunable KE, variable "building blocks", ability to deposit through background gases) rapidly increase the complexity of the growth process, making optimization steps nearly impossible to predict *via* human intuition, and an ideal process for autonomous experimentation and development. Here we demonstrate that Gaussian process regression and Bayesian optimization[25] in Auto-PLD can autonomously and efficiently discover growth window and process-property relationships within a broad, 4D parameter space, while automation improvements increase the throughput at least ~10x relative to traditional PLD workflows. Using the Auto-PLD and approach, the autonomous synthesis of any material that can be grown by PLD should be enabled, providing a great opportunity to accelerate synthesis science.

**Results**

**Autonomous Synthesis Workflows with Pulsed Laser Deposition**

The primary goal of autonomous synthesis is not only to accelerate growth experiments *via* automation but also to gain a greater depth of understanding of the synthesis process than a human experimentalist could achieve by undertaking the same set of experiments, thereby reducing the overall number of experiments that need to be performed in order to reach a conclusion. For instance, experimentalists with years of domain knowledge can find the growth window for a material without the aid of AI by exploring a tiny fraction of the parameter space (varying 1 or 2 parameters) but learn very little about the global relationship between mulitple experimental parameters and sample properties. As we will show later, autonomous experiments can find the growth window for a material and also learn the approximate process-property relationship for the entire parameter space within in a smaller number of samples than would be possible for a human experimentalist alone. Once learned, the full process-property relationship can be mapped to physically meaningful values such as chemical potential to directly test and provide feedback to theoretical predictions of phase diagrams or stability. Another key advantage of autonomous synthesis is the ability to monitor multiple feedback-streams and make real-time parameter adjustments during growth that would be difficult or impossible for a human to do. In these ways, autonomous synthesis platforms can achieve accelerated understanding and control in synthesis science.

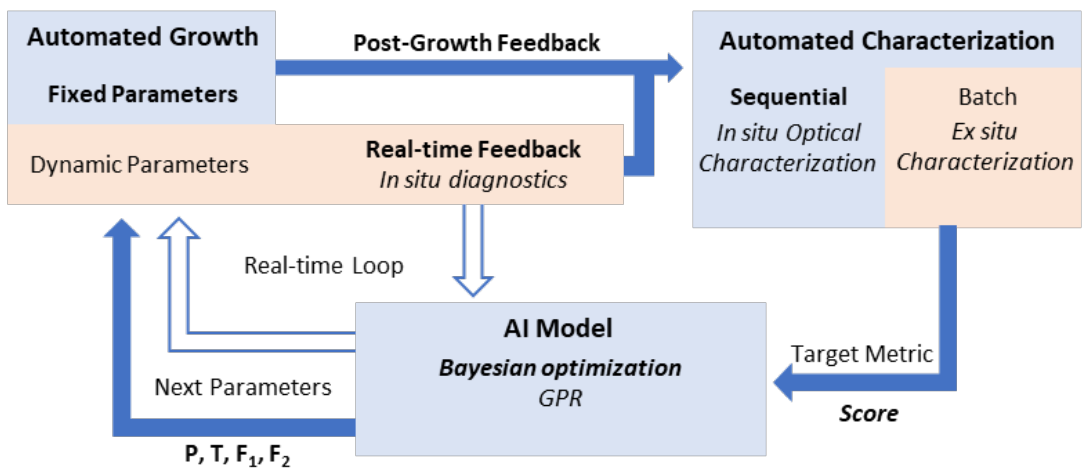

**Figure 1. Workflows for autonomous PLD synthesis.** Samples are synthesized with a set of parameters and feedback is collected through automated material characterizations. A target metric is derived from the feedback and used to train an AI model to predict a new set of synthesis conditions. The feedback/AI model can be either post-growth or in real-time depending on the goals of the autonomous experiment. The current study uses real-time feedback to augment the growth routine and sequential post-growth feedback based on *in situ* Raman spectroscopy (*score*) to train a Gaussian process Bayesian optimization model that suggests values for growth parameters ($P, T, F_1, F_2$) for the next experiment.

Workflows for autonomous synthesis experiments are a closed-loop, using cyclic steps of automated synthesis, automated characterization, and intelligent decision making with AI. **Figure 1** shows a schematic of the possible autonomous PLD synthesis workflows. In general, a sample is synthesized with specified parameters, feedback (target metric) is collected through material characterizations, and an AI model predicts new values for the chosen set of growth parameters that should improve the result or decrease uncertainty in the parameter space. The most straightforward autonomous experiment uses feedback that is acquired post-growth such as a Raman spectrum, x-ray diffractogram, or electrical transport measurement. The AI model is then updated with this feedback either sequentially (sample-by-sample) or in batches (*e.g.*, 10 samples at a time) to predict the next set of parameters. Real-time feedback acquired during growth such as the evolution of a Raman spectrum[23], reflectivity[26], or a RHEED pattern[27] can also be monitored to either augment the growth routine or act as feedback for a real-time AI model to *dynamically* adjust parameters *during* synthesis.

The workflow of the autonomous experiment in the present study uses a combination of post-growth and real-time feedback. Real-time laser reflectivity is used to control film thickness while post-growth *in situ* Raman spectroscopy is used as feedback to a Gaussian process Bayesian optimization model which attempts to maximize the crystallinity of ultrathin $WSe_2$ films. In **Figure 1,** this is the workflow is highlighted by the solid blue arrows and boxes: fixed growth parameters, real-time feedback (without dynamic parameter updates), and sequential post-growth characterization. We are not currently using the real-time loop defined by the red boxes and open arrows which would be used to dynamically change growth parameters during synthesis or any *ex situ* characterization for AI model feedback. In the sections below we detail the platform design that enables autonomous PLD synthesis, the Bayesian optimization algorithm and target metric, and show the results of our autonomous growth experiment.

**Automated Pulsed Laser Deposition Platform**

The Auto-PLD platform is designed for fully automated synthesis of lab-scale samples with correlated *in situ* diagnostics of the plasma plumes and *in situ* optical characterization of the growing film. **Figure 2a** shows a diagram of the Auto-PLD system (**Figure S1** gives additional information.) The system has two targets which can be ablated simultaneously or separately using two, individually attenuated laser beams to permit mixing the flux from two materials during growth to account for volatile element losses, dope materials, or create layered films. The ability to mix two plumes with co-ablation enables the automation of target composition experiments without the need to synthesize numerous PLD targets, which is of particular interest for materials with volatile elements like $WSe_2$. Plume conditions can be monitored by gas-phase plasma diagnostics, including two translatable ion probes which yield ion currents vs. position along the normal to each target, gated-ICCD photography of the visible plasma plume intensity and spatial propagation, and gated-ICCD spectroscopy that can be selected at different locations (*e.g.,* just above the substrate) and times, in order to evaluate reproducibility or correlate plume conditions to diagnostics of the sample. To address the necessity of high-throughput methodology for autonomous synthesis, we developed a laser-heated substrate exchange wheel that allows for the sequential growth of up to 10 samples without the need for human intervention. The active substrate is placed in line with the laser heater while the remaining 9 substrates are shielded from the plumes. **Figure 2b** shows the substrate wheel with the shield partially cut away to display the inactive substrates. The substrates are mounted on standard flag-style sample plates to allow for easy transfer to other vacuum-based techniques such as XPS, STM, ARPES, etc. Additionally, the Auto-PLD is integrated into a laboratory network ecosystem that automates sample data/metadata structuring and data transfer to and from cloud resources (**Figure S2** and **Note S1**).

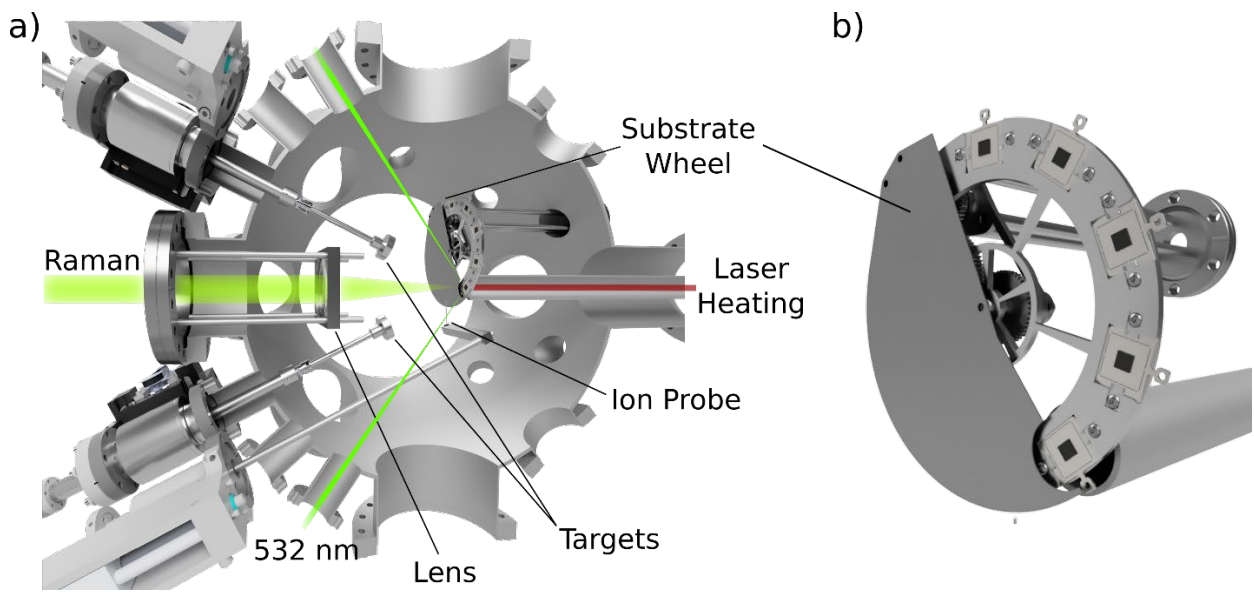

**Figure 2. The Auto-PLD system employs multiple *in situ* diagnostics and a multi-substrate exchange wheel to enable autonomous synthesis.** a) The side view cross section of the chamber (Y-Z plane) shows the geometry of the 2 off-axis PLD targets and one of the translatable ion probes, 532 nm laser path for *in situ* Raman spectroscopy and photoluminescence, collection lens normal to the substrate for Raman/PL, and the 10-substrate exchange wheel with backside 976 nm laser heating and pyrometry. Optical reflectivity from the substrate (laser or white light) is performed in the orthogonal (X-Z) plane (see Supporting Information). Large windows allow gas-phase plasma diagnostics (gated ICCD imaging and spectroscopy). b) The exchange wheel holds up to 10 substrates and is used to rotate the active substrate into place for backside laser heating while the remaining substrates are shielded from the plumes (shield partially cut away for clarity).

Various optical measurements or diagnostics of the sample are possible. Spectroscopic information is collected from the substrate through an internal lens and projected out of the chamber to a spectrometer. More generally, the chamber can be equipped with many optical characterizations and processing such as Raman, second harmonic generation (SHG), photoluminescence (PL), white light reflectance, or lasers for processing or fast sintering, which make the Auto-PLD system a highly flexible synthesis and processing platform. Automated spectroscopy routines also allow data acquisition during growth which enables us to watch the film evolve in real time with time resolution ranging from milliseconds to seconds. See the Methods section for details on the Raman spectroscopy and laser reflectivity that were used in this study.

In this study, we demonstrate the autonomous exploration of a broad (4D) synthesis parameter space with a minimal number of experiments using *in situ* optical reflectivity as real-time feedback for thickness control, and *in situ* Raman spectroscopy for post-growth characterization feedback. Plasma diagnostics (ICCD-imaging sequences, and ion probe currents) are automatically collected for each run and are currently used only for correlation with the final results of the autonomous workflow developed here. However, with the simulation and modeling of the plume dynamics or growth kinetics, plume imaging or real-time Raman diagnostics for growth kinetics and metastable state tracking could be used within the autonomous routines as target metrics or control parameters.

**Bayesian Optimization and Defining a Target Metric**

Bayesian optimization (BO) has been applied to various complex problems with parameter spaces that are discontinuous[28], discrete (user preference)[29,30], or high dimensional[31,32]. Likewise, BO has been widely applied to accelerate material discoveries with rapid exploration of control parameter spaces[33–35] and develop autonomous platforms[36,37]. Here, motivated by accelerating the understanding of material synthesis, we implement BO to develop an autonomous PLD synthesis platform to rapidly discover process-property relationships in large parameter spaces.

Bayesian optimization is an experimental design strategy which is typically used to efficiently find a global optimum of an unknown objective function when acquiring samples from the objective is difficult to obtain, expensive, or time consuming. In this context, sampling a point of the objective function represents synthesizing and characterizing a thin film, processes which are both expensive and time consuming. The idea behind BO follows Bayes' theory, where given a few samples of the parameter space, the data is fitted with a cheaper posterior surrogate model, generally a Gaussian process regression model (GPR)[38], to approximate the functional relationship between the objective and its parameters. Once the GPR measures the mean and the uncertainty of all the unexplored regions in the parameter space the choice of the future expensive experiments is guided by an adaptive sampling strategy through maximizing an acquisition function. Several acquisition functions, such as Probability of Improvement (PI), Expected Improvement (EI), and Confidence Bound criteria (CB) have been developed with different trade-offs between exploration and exploitation[39–42]. Here, maximizing the acquisition function is analogous to optimizing the expensive and/or black-box objective function. The detailed mathematical formulation of the BO is provided in the Methods Section.

Our optimization target metric (objective function) in the current experiment is based on Raman spectroscopy. After each sample growth, we use *in situ* Raman spectroscopy to evaluate film crystallinity. Raman scattering in ultrathin transition metal dichalcogenide (TMD) materials is sensitive to the number of layers, defects, strain, stoichiometry, and manifests in the measured Raman spectrum as changes in relative mode intensities, mode frequency, increased linewidth, and appearance of defect-related modes[43–45]. As such, the crystallinity or "quality" of a TMD can be evaluated through Raman spectroscopy as a first approximation. For $WSe_2$, the Raman spectrum is dominated by a single peak, comprised of the $E_{2g}$ and $A_{1g}$ modes located near 251 cm$^{-1}$ and 247 cm$^{-1}$ in bulk, which will be referred to as the $E_{2g}+A_{1g}$ peak. Highly defective $WSe_2$ shows a reduction in the relative intensity and linewidth broadening of the $E_{2g}+A_{1g}$ peak[46]. It has been shown for CVD samples that the $E_{2g}+A_{1g}$ intensity for monolayer $WSe_2$ is maximized and the full width at half maximum (FWHM) is minimized for optimal samples[47]. Considering this, we calculate a *score* for each Raman spectrum by taking the ratio of the peak prominence and the FWHM of the $E_{2g}+A_{1g}$ peak, where the prominence and FWHM are determined by a multi-Lorentzian curve fitting routine (**Figure S3** shows an example of the fitting routine.) The peak prominence is used, rather than intensity, to make the calculated score less sensitive to the absence of a peak in the spectrum (i.e. the fixed peak fitting routine will always return values, even if a real peak is not present in the data.) We calculate the peak prominence from the "peak_prominences" function in the Python "SciPy" library package[48] from the slope, calculated at each discretized points of the $E_{2g}+A_{1g}$ peak (using "find_peaks" function) of the multi-Lorentzian fitted Raman spectrum. The FWHM measures the sharpness of the peak where low FWHM value maps high sharpness. Finally, we maximize this score objective function as a function of growth parameters in the BO framework.

In addition to the crystallinity, the Raman intensity is generally related to the fractional surface coverage of an ultrathin film within the laser excitation area (Raman intensity is a function of scattering volume[49]). So, if the nominal thickness of a series of ultrathin films is fixed to be approximately 1 monolayer, the *score* metric simultaneously optimizes the coverage of the film which accounts for incomplete coalescence. As such, the *score* metric we are using with the $E_{2g}+A_{1g}$ Raman peak in this study should simultaneously optimize the surface coverage and the crystallinity of the nominally monolayer films.

**Autonomous PLD Synthesis Experiment**

For the autonomous growth of $WSe_2$ on $Si/SiO_2$ substrates, we chose to vary 4 parameters that play a major role in the deposition: laser fluence on the $WSe_2$ target $F_1$, laser fluence on the Se target $F_2$, background pressure P, and substrate temperature T (see Methods for details). All other parameters are fixed except for the number of laser pulses. Because all 4 of the varied parameters will affect the growth rate, we grow each sample to the same nominal thickness instead of fixing the number of laser pulses. To do this, we monitor the optical contrast in real-time using laser reflectivity and continuously add pulses until the contrast for a nominal monolayer of $WSe_2$ is reached (see Methods). Although the high supersaturation inherent to PLD leads to 2$^{nd}$ and 3$^{rd}$ layer nucleation before a full monolayer can coalesce on SiO$_2$ it is a reasonable optical metric to approximate film thickness for this sub 5 nm thickness regime[26]. For a detailed description of the workflow, see Supporting Information **Note S2**.

Before using the BO for parameter suggestions, we synthesized 10 samples using the same automated growth routine and characterization but with growth conditions chosen by the experimenter. These 10 samples were used to initialize the BO, and then 115 samples were grown autonomously. The 125 samples required 114 hours of instrument time, averaging 55 minutes to grow, characterize, and train the BO per sample. Reloading the substrate wheel, cleaning the laser window, and pumping the chamber down took ~2 hours per batch (26 hours total). The total time to complete all 125 samples was ~140 hours. Thus, the Auto-PLD produced 21 samples/24 hour period during this experimental campaign. A human experimentalist using a traditional PLD workflow can grow and characterize 2-3 samples/24 hour period with the same chamber, without the sample wheel, the same vacuum pumping and substrate heating times, and including *in situ* Raman which is not typical for PLD systems. Thus, we see a ~10x increase in throughput over traditional PLD by using the autonomous system. In the Auto-PLD routine, many of the trials reached the deposition cut off time of 45 minutes in conditions where no material was depositing, which greatly increases the average time per sample. Nearer to optimal conditions, the throughput will increase due to faster average growth times. **Figures S4.1-3** show examples of the *in situ* diagnostics collected during growth for several samples.

**Figure 3** shows the results of the 115 BO steps. The 4D surrogate function is projected onto each 2D parameter plane by averaging over the other 2 parameters and the scattered points represent the samples that were grown where darker red points indicate a higher score. The surrogate provides an approximation for the process-property relationship between P, T, $F_1$, $F_2$, and the score which is clearly seen to be non-linear in this case. To minimize the number of samples grown at the boundaries of the parameter space and to favor exploration, we used a periodic kernel for the first 105 BO steps because the periodic kernel is more likely to avoid exploring near the edges of the parameter space during the early iterations. The evolution of the BO surrogate function and variance is shown in **Figure S5.1-2**. Since a periodic relationship between P, T, $F_1$, and $F_2$ is highly unlikely to represent the real physical behavior of this system, we switched to an RBF kernel for the final 10 samples (later iterations when the central part of the space has been sufficiently explored), and the surrogate function as calculated with the RBF kernel after the last step is shown in **Figure 3a-f**.

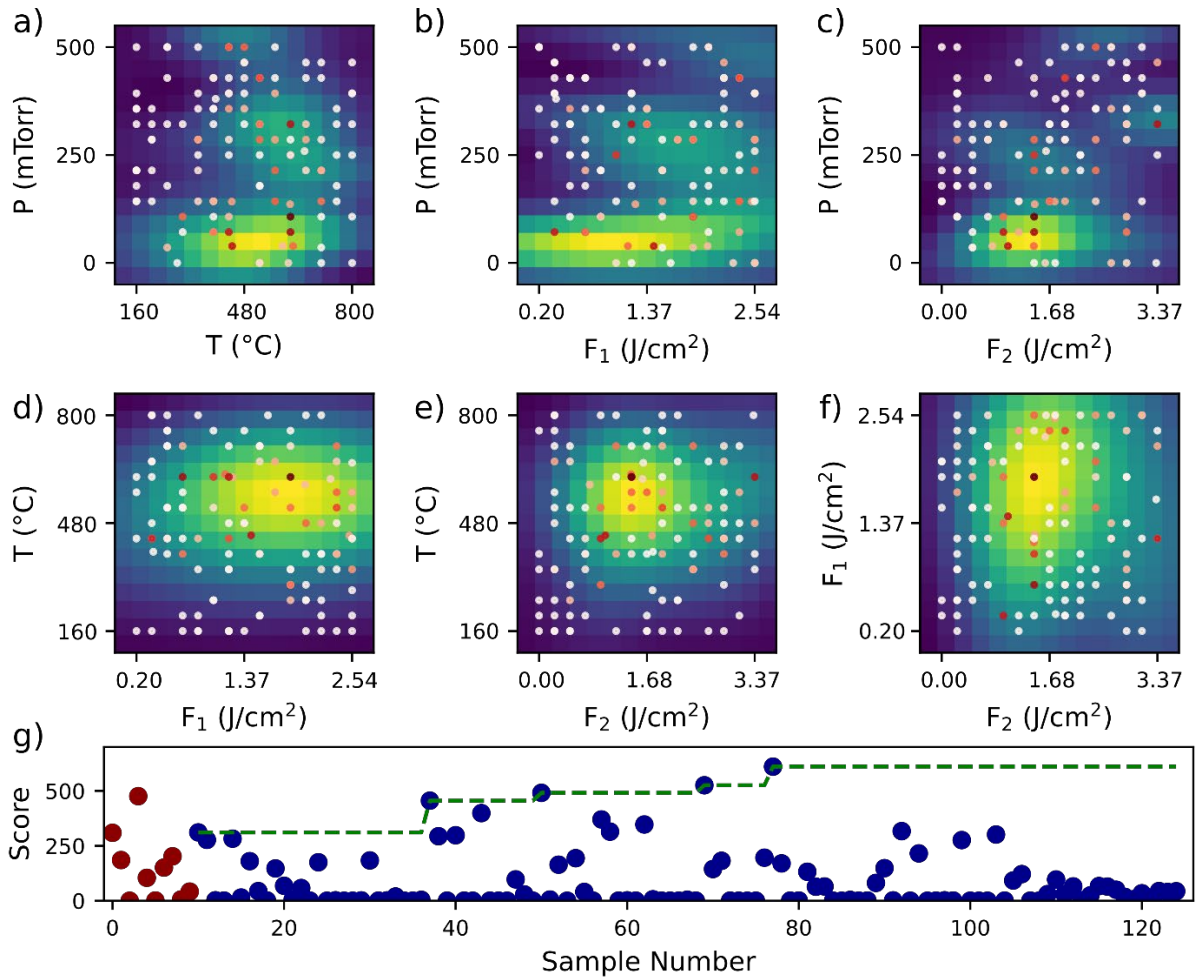

**Figure 3. 2D representations of the GPR surrogate function after 125 growth steps.** The 4D GPR surrogate function is represented in each 2D parameter plane and shows the predicted process-property relationship between P, T, $F_1$, $F_2$, and the *score*. The P planes a-c) show a localized global optimum below 100 mTorr with a 2nd local optimum between 200-350 mTorr. The T planes d-e) and the F plane f) show a less localized optima relative to the P planes. The best predicted parameters from this model are located at P = 71.4 mTorr, T = 526 °C, $F_1$ = 1.20 J/cm$^2$, and $F_2$ = 1.44 J/cm$^2$. g) The score vs. sample number indicates a gradual improvement in sample quality over time and indicates that the model is still exploring the space to reduce uncertainty rather than exploiting it to converge on the global maximum. The dashed green line in g) tracks the highest score found by the BO.

**Figure 3g** shows the *score* vs sample number which indicates a gradual improvement in the sample quality over time and that the BO is exploring to reduce the uncertainty in the parameter space, rather than exploiting it to converge on the global optimum, which manifests as suggesting conditions with high uncertainty (which generally produce poor quality samples due to the narrow window in P).

The global optimum predicted by the BO is located at P = 71.4 mTorr, T = 526 °C, $F_1$ = 1.20 J/cm$^2$, and $F_2$ = 1.44 J/cm$^2$. The region around the global surrogate optimum, i.e. the growth window, is highly localized in the P planes (**Figure 3a-c**) and appears to have another local maximum between 200-350 mTorr. By comparison, the growth windows in the T (**Figure 3d-e**) and fluence (**Figure 3f**) planes are broader and feature a single optimum. The overall growth window discovered by the Auto-PLD experiment agrees with what other PLD studies of TMD growth manually found. Typical parameters from literature for WSe$_2$ are T = 500°C, $F_1$ = 1 J/cm$^2$, an Ar background P = 100 mTorr, with a single, Se-compensated target[50,51] which act as validation of the AI model's results. WSe$_2$ films synthesized within the growth window also show the millimeter-scale uniformity and < 1 nm surface roughness (**Figure S6**) that is typical of PLD-grown TMD films[52,53].

It is worth noting that the BO has not converged with 125 samples and we did not expect convergence in a 4D BO with a small percentage of the space having been sampled. However, the BO provides a reasonable qualitative relationship between the growth parameters and the sample quality despite having sampled only 0.25% of the parameter space (total number of possible parameter combinations is 50,625). As a simple test of how well the surrogate can predict the score, we synthesized 10 samples with different T at fixed P = 86.2 mTorr, $F_1$ = 1.08 J/cm$^2$, and $F_2$ = 1.39 J/cm$^2$ without updating the BO. None of these 10 parameter combinations were present in the training data and there is only one sample grown nearby at P = 250 mTorr, T = 617 °C, $F_1$ = 1.03 J/cm$^2$ $F_2$ = 1.44 J/cm$^2$. **Figure S7** shows how the BO effectively reproduces the rise and fall of the score with increasing T, even along a line that has never been sampled. This indicates that the GPR surrogate model has acquired enough data (0.25% of the space) to effectively approximate the relationship between the growth parameters and the sample score.

These results indicate that both Se compensation and a background gas are required to grow high quality WSe$_2$ samples. Se loss during growth is likely caused by incongruent evaporation of W and Se from the stoichiometric WSe$_2$ target and differences in the sticking coefficient of each in different conditions. The P dependence is more complex because increasing P changes the plume expansion dynamics, plume-plume interactions, deposition rate, and the angular and kinetic energy distributions of plume species. We attribute the increase in quality from 0-100 mTorr to the decreased kinetic energy of plume species. Beyond 100 mTorr, the Se plume begins to deflect the slow component of the WSe$_2$ away from the substrate, see the ICCD image sequences in **Figure S4.1-3**. The deflection of the slow component along with the expectation of increased nanoparticle formation at higher pressures suggests the delivery of fundamentally different building blocks for film growth relative to the lower pressure regime. Thus, the local maxima in P between 200-350 mTorr may represent a growth regime dominated by nanoparticles with a different fundamental growth mechanism than the lower pressure region.

**Discussion**

We demonstrated how automation and machine learning can be incorporated with PLD to autonomously discover growth windows in a 4D parameter space spanning background pressure, substrate temperature, and laser fluence on two PLD targets. The large parameter space was autonomously explored with Bayesian optimization and discovered two distinct growth windows and the process-property relationship in the chosen parameter space for ultrathin WSe$_2$ films. Full automation of synthesis, sample exchange, and characterization showed at least a 10x increase in throughput compared to traditional PLD workflows while Bayesian optimization with Gaussian process regression provided a predictive model for film quality after sampling only 0.25% of the parameter space. Based on the *in situ* plume imaging and ion probe diagnostics, we attribute different growth mechanisms to the two observed growth windows which are dominated by different plume expansion dynamics, plume-plume interactions, deposition rate, and angular and kinetic energy distributions of plume species.

Although we varied P, T, $F_1$, and $F_2$ in the current experiment, more parameters such as repetition rate, timing between ablation of the two targets, pulse number sequences, background gas compositions, laser spot size, wavelength, and pulse duration, post-growth annealing temperatures and times can be included in the workflow for further exploration of the growth window and to locate more precise optimum synthesis parameters. The real parameter space for PLD (PVD) synthesis is vast and underscores the need for well-designed automated experiments and the necessity to develop AI/ML techniques to address these large parameter spaces with sparse sampling. From an AI standpoint, the Auto-PLD enables the development of many other AI-driven experiments using models like multi-objective Bayesian optimization, reinforcement learning with real-time diagnostic feedback, hybrid AI-human collaborative experiments, and physics-informed ML models. From an experimental perspective, the atypical, dual-target PLD chamber design along with customizable *in situ* spectroscopy and ample optical access to the sample enables unique growth experiments incorporating co-ablation of two targets or laser processing, crystallization studies with amorphous layers deposited *in situ*, and plume diagnostic/spectroscopy studies which can all be done in an automated, sequential fashion with minimal human intervention. Further, real-time diagnostic measurements like laser reflectivity or Raman spectral evolution can be used to learn how the dynamics of growth affect the outcome, which can lead to directed or precision growth with a mechanistic understanding[23]. In the future, we plan to incorporate details of the gas phase diagnostics like the plume KE and presence of different plume components[54] to inform the AI decision making by correlating the plume properties to the outcome of the film growth and to aid in human task of choosing the growth parameter space i.e., creating a well-designed experiment. By incorporating these gas phase correlations, we expect that autonomous PLD synthesis will be more efficient in exploration, achieve faster convergence to the global optimum, and can reveal how processing parameters are linked to fundamental growth mechanisms.

The versatility of PLD enables the synthesis/study of a vast number of materials systems and the Auto-PLD can be used to explore autonomous experimentation and accelerate discovery or optimization with any material that can be grown by PLD. Even for properties that are not optically accessible, automated *ex situ* characterizations can be used with a batch update approach, making the autonomous workflows described here universal for PVD synthesis techniques. Growth platforms such as this one allows researchers to focus on solving scientific problems, rather than being deposition machines themselves, and increase the rate of discovery, optimization, and understanding in the synthesis of materials.

**Methods**

**Substrate Preparation**

The 5x5 mm substrates used for the autonomous growth experiment were all diced from the same 3-inch 90 nm $SiO_2$/Si wafer (University Wafer, ID: 3595, Dry Thermal Oxide). Substrates were sonicated in acetone, methanol, and isopropyl alcohol for 5 minutes each and blown dry with $N_2$. Silver paste was used to bond the substrates to the sample plates and were baked on a hot plate for 20 minutes at 120°C before loading into the chamber.

**Pulsed Laser Deposition and Plume Diagnostics**

Two PLD targets made of WSe$_2$ (99.8%, Plasmaterials) and Se (99.999%, Plasmaterials, Inc.) were ablated simultaneously using a KrF excimer laser (Coherent LPX 305F, 248 nm, 25 ns, 2 Hz rep rate) with target to substrate distances of 5 cm and 10 cm, respectively. Each target is offset from the substrate normal by 25°. The KrF beam is split into two beamlets which pass through separate, motorized energy attenuators and rectangular apertures. The apertures are imaged onto the targets using a projection beamline. The beam spot sizes on each target were 0.0256 cm$^2$ (10x10 mm input aperture) and 0.0410 cm$^2$ (20x15 mm input aperture) for WSe$_2$ and Se, respectively. The laser fluence for each target was adjusted with the attenuators for each beamlet while maintaining the same spot size. The Ar background pressure (99.9999%, 5 sccm) was regulated with a throttle valve and a mass flow controller. The base pressure for each set of 10 samples was < 5x10$^{-6}$ Torr. The substrates were heated with a remote 976 nm, 140 W laser from the backside of the Inconel sample plates and the temperature was measured with a backside pyrometer to within ±1°C.

Sequences of 50 ICCD images were collected during each deposition with delay times from 1-150 μs and the gate time set to 10% of the delay time for each image. The ICCD camera (Princeton Instruments, PI-MAX 4) was positioned 80 cm away from the center of the plume and used an f2.8 camera lens (105 mm micro-NIKKOR). Ion probe waveforms were measured using a biased wire placed next to the substrate (-40V bias, 4mm long, 0.4mm diameter) and an oscilloscope (1 GHz, Tektronix MSO64) with a 50 Ω feed-through BNC resistor. The target-probe distances were 4.5 cm and 9.5 cm for the WSe$_2$ and Se targets, respectively. The ICCD camera and the oscilloscope were both triggered by a photodiode monitoring the KrF laser pulses.

***In situ* Raman Spectroscopy**

A 532 nm continuous wave laser (Cobolt Samba, 1 W max, Huber Photonics) was used to excite the samples at 55° angle of incidence. The laser spot on the samples was elliptical (0.7x0.6mm) with the major axis aligned with the spectrometer slit. Raman scattered light was collected at normal incidence and imaged onto the spectrometer slit with a 75 mm dia., $f$ = 350 mm spherical lens mounted inside the chamber. A long-pass edge filter (Semrock, RazorEdge) was used to filter the laser line. The spectrometer (Princeton Instruments, Isoplane SCT 320) was coupled with a CCD detector (Princeton Instruments, PIXIS 256e) and used a 2400 groves/mm holographic grating. The laser power and CCD exposure times for all samples and reference spectra were 300 mW and 15s exposure, 4 averages.

**Real-time Laser Reflectivity**

Reflectivity was monitored using a randomly polarized, stabilized HeNe laser (632.8 nm, 1.2 mW, Thorlabs, Inc, HRS015B) with an incident angle of 32.5°. The beam was randomly polarized using a liquid crystal polymer depolarizer (Thorlabs, Inc., DPP25-B). Reflected intensity was measured through a laser line filter (Thorlabs, Inc. FL632.8-1) using a photodiode (Thorlabs, Inc., SM1PD1B) and a source measure unit (Keithley 2450 SMU). Reflectivity was acquired for 10 seconds prior to the start of deposition to determine an average initial value $I_0$, then the optical contrast $C(t)$ was calculated and monitored in real-time using the equation $C(t) = (I(t) - I_0)/I_0$ where $I(t)$ is the photodiode current at time $t$.

The optical contrast of the Si/SiO$_2$/WSe$_2$ layer stack was modeled by calculating the Fresnel reflection coefficients using the recursive relations for successive layers, demonstrated in a previous work[26]. The randomly polarized reflection coefficient was calculated as the average of the s- and p-polarization coefficients. Refractive indices of 3.87-0.016$i$ and 1.47 were used for Si and SiO$_2$[55] and 4.42-0.60$i$ was used for WSe$_2$[56]. With these parameters, 1 monolayer of WSe2 on 90 nm SiO$_2$/Si gives a contrast of -0.364, which was the target contrast to trigger the end of each deposition. Temperature dependent changes in the refractive indices were neglected for this study. Uncertainty in the angle of incidence (±2.5°) causes error in $C(t)$ below the measurement noise. A Python module for calculating the contrast of the layer stack and a Jupyter notebook are provided online at https://github.com/sumner-harris/Fresnel-TMD-Contrast.

*Ex Situ* **Characterization**

Raman mapping was used to assess the millimeter scale uniformity of the WSe$_2$ films using a custom-built spectroscopy microscope with a 10x objective lens, 320 µW of 532 nm laser excitation, and a 1800 grooves/mm grating with 20s exposure. The map was collected over a 1x1 mm$^2$ area with a step size of 0.1 mm.

Atomic force microscopy (AFM) was used to measure the roughness of the films and was performed with a Bruker Dimension Icon AFM in tapping mode with a Si probe (TESPA-V2, 7 nm tip radius, 37 N/m spring constant.)

**Bayesian Optimization**

The parameter space for BO was discretized to give 15 evenly spaced, experimentally significant values for each parameter (e.g. a substrate temperature of 401°C is not expected to give a result that is significantly different from 400°C) giving the space a size of 50,625 points. We used periodic and RBF kernel functions and the Adam optimizer with learning rate of 0.1 to fit and train the GPR model and the Expected Improvement acquisition function for adaptive sampling in BO.

The general form of the GPM is as follows:

$$y(x) = x^T \beta + z(x) \tag{1}$$

where $x^T\beta$ is the Polynomial Regression model. The polynomial regression model captures the global trend of the data. $z(x)$ is a realization of a correlated Gaussian Process with mean $E[z(x)]$ and covariance $cov(x^i, x^j)$ functions defined as follows:

$$z(x) \sim GP\left(E[z(x)], cov(x^i, x^j)\right); \tag{2}$$

$$E[z(x)] = 0, cov(x^i, x^j) = \sigma^2 R(x^i, x^j) \tag{3}$$

$$R(x^i, x^j) = R_p(x^i, x^j) = \exp\left(-2 * \sum_{m=1}^{d} \frac{\sin^2\left(\pi|x_m^i - x_m^j|/p\right)}{\theta_m^2}\right); \tag{4}$$

$$R(x^i, x^j) = R_r(x^i, x^j) = \exp\left(-0.5 * \sum_{m=1}^{d} \frac{\left(x_m^i - x_m^j\right)^2}{\theta_m^2}\right); \tag{5}$$

$$\theta_m = (\theta_1, \theta_2, \ldots, \theta_d)$$

where $\sigma^2$ is the overall variance parameter, $\theta_m$ is the correlation length scale parameter in dimension $m$ of $d$ dimension of $x$, $p$ is the parameter period which determines the distance between repetitions of the function. These are termed as the hyper-parameters of GPR model. $R_p(x^i, x^j)$ and $R_r(x^i, x^j)$ are the spatial correlation function with periodic and RBF kernel respectively.

We maximize the Expected improvement acquisition function $\underset{x_i \in \bar{\bar{X}}}{\operatorname{argmax}} u(\bar{\bar{Y}}(\bar{\bar{X}})|\Delta_k)$ as

$$u(\bar{\bar{y}}(\bar{\bar{x}})|\Delta_k) = EI(\bar{\bar{y}}(\bar{\bar{x}})) =$$

$$\begin{cases} (\mu(\bar{\bar{y}}(\bar{\bar{x}})) - y(x^+) - \xi) * \Phi(Z, 0, 1) + \sigma(\bar{\bar{y}}(\bar{\bar{x}})) * \phi(Z) & if\ \sigma(\bar{\bar{y}}(\bar{\bar{x}})) > 0 \\ 0\ if\ \sigma(\bar{\bar{y}}(\bar{\bar{x}})) = 0 \end{cases} \tag{6}$$

$$Z = \begin{cases} \frac{\mu(\bar{\bar{y}}(\bar{\bar{x}})) - y(x^+) - \xi}{\sigma(\bar{\bar{y}}(\bar{\bar{x}}))} & if\ \sigma(\bar{\bar{y}}(\bar{\bar{x}})) > 0 \\ 0\ if\ \sigma(\bar{\bar{y}}(\bar{\bar{x}})) = 0 \end{cases} \tag{7}$$

where $u(\overline{\overline{Y}}(\overline{\overline{X}})|\Delta_k)$ is the vector of acquisition function values of all the non-sampled inputs (where experiments are not conducted) $\overline{\overline{Y}}(\overline{\overline{X}})$ given the posterior model at a given BO iteration k, $y(x^+)$ is the current maximum value among all the sampled inputs until the current stage which is at $x = x^+$; $\mu(\bar{\bar{y}})$ and $\sigma^2(\bar{\bar{y}})$ are the predicted mean and MSE from GPR for the inputs $\bar{\bar{x}} \in \overline{\overline{X}}$; $\Phi(.)$ is the cdf; $\phi(.)$ is the pdf; $\xi \geq 0$ is a small value which is set as 0.01.

## Acknowledgements

This work was supported by the U.S. Department of Energy, Office of Science, Basic Energy Sciences, Materials Sciences and Engineering Division. The development and deployment of software for the autonomous synthesis routine and the *ex situ* characterization was supported by the Center for Nanophase Materials Sciences (CNMS), which is a US Department of Energy, Office of Science User Facility at Oak Ridge National Laboratory.

## References

(1) Pyzer-Knapp, E. O.; Pitera, J. W.; Staar, P. W. J.; Takeda, S.; Laino, T.; Sanders, D. P.; Sexton, J.; Smith, J. R.; Curioni, A. Accelerating Materials Discovery Using Artificial Intelligence, High Performance Computing and Robotics. *NPJ Comput Mater* **2022**, *8* (1), 84. https://doi.org/10.1038/s41524-022-00765-z.

(2) de Yoreo, J.; Mandrus, D.; Soderholm, L.; Forbes, T.; Kanatzidis, M.; Erlebacher, J.; Laskin, J.; Wiesner, U.; Xu, T.; Billinge, S.; Tolbert, S.; Zaworotko, M.; Galli, G.; Chan, J.; Mitchell, J.; Horton, L.; Kini, A.; Gersten, B.; Maracas, G.; Miranda, R.; Pechan, M.; Runkles, K. *Basic Research Needs Workshop on Synthesis Science for Energy Relevant*

*Technology*; U.S. Department of Energy, Office of Science: Washington, DC, 2016. https://doi.org/10.2172/1616513. (accessed on June 15, 2023).

(3) Abolhasani, M.; Kumacheva, E. The Rise of Self-Driving Labs in Chemical and Materials Sciences. *Nature Synthesis* **2023**, 2, 483-492. https://doi.org/10.1038/s44160-022-00231-0.

(4) Reinhardt, E.; Salaheldin, A. M.; Distaso, M.; Segets, D.; Peukert, W. Rapid Characterization and Parameter Space Exploration of Perovskites Using an Automated Routine. *ACS Comb Sci* **2020**, *22* (1), 6–17. https://doi.org/10.1021/acscombsci.9b00068.

(5) Szymanski, N. J.; Zeng, Y.; Huo, H.; Bartel, C. J.; Kim, H.; Ceder, G. Toward Autonomous Design and Synthesis of Novel Inorganic Materials. *Mater Horiz* **2021**, *8* (8), 2169–2198. https://doi.org/10.1039/D1MH00495F.

(6) Li, Y.; Xia, L.; Fan, Y.; Wang, Q.; Hu, M. Recent Advances in Autonomous Synthesis of Materials. *ChemPhysMater* **2022**, *1* (2), 77–85. https://doi.org/10.1016/j.chphma.2021.10.002.

(7) Epps, R. W.; Felton, K. C.; Coley, C. W.; Abolhasani, M. Automated Microfluidic Platform for Systematic Studies of Colloidal Perovskite Nanocrystals: Towards Continuous Nano-Manufacturing. *Lab Chip* **2017**, *17* (23), 4040–4047. https://doi.org/10.1039/C7LC00884H.

(8) Volk, A. A.; Epps, R. W.; Yonemoto, D. T.; Masters, B. S.; Castellano, F. N.; Reyes, K. G.; Abolhasani, M. AlphaFlow: Autonomous Discovery and Optimization of Multi-Step Chemistry Using a Self-Driven Fluidic Lab Guided by Reinforcement Learning. *Nat Commun* **2023**, *14* (1), 1403. https://doi.org/10.1038/s41467-023-37139-y.

(9) Burger, B.; Maffettone, P. M.; Gusev, V. V; Aitchison, C. M.; Bai, Y.; Wang, X.; Li, X.; Alston, B. M.; Li, B.; Clowes, R.; Rankin, N.; Harris, B.; Sprick, R. S.; Cooper, A. I. A Mobile Robotic Chemist. *Nature* **2020**, *583* (7815), 237–241. https://doi.org/10.1038/s41586-020-2442-2.

(10) MacLeod, B. P.; Parlane, F. G. L.; Morrissey, T. D.; Häse, F.; Roch, L. M.; Dettelbach, K. E.; Moreira, R.; Yunker, L. P. E.; Rooney, M. B.; Deeth, J. R.; Lai, V.; Ng, G. J.; Situ, H.; Zhang, R. H.; Elliott, M. S.; Haley, T. H.; Dvorak, D. J.; Aspuru-Guzik, A.; Hein, J. E.; Berlinguette, C. P. Self-Driving Laboratory for Accelerated Discovery of Thin-Film Materials. *Sci Adv* **2022**, *6* (20), eaaz8867. https://doi.org/10.1126/sciadv.aaz8867.

(11) Pereira, D. A.; Williams, J. A. Origin and Evolution of High Throughput Screening. *Br J Pharmacol* **2007**, *152* (1), 53–61. https://doi.org/10.1038/sj.bjp.0707373.

(12) Macarron, R.; Banks, M. N.; Bojanic, D.; Burns, D. J.; Cirovic, D. A.; Garyantes, T.; Green, D. V. S.; Hertzberg, R. P.; Janzen, W. P.; Paslay, J. W.; Schopfer, U.; Sittampalam, G. S. Impact of High-Throughput Screening in Biomedical Research. *Nat Rev Drug Discov* **2011**, *10* (3), 188–195. https://doi.org/10.1038/nrd3368.

(13) Nikolaev, P.; Hooper, D.; Perea-López, N.; Terrones, M.; Maruyama, B. Discovery of Wall-Selective Carbon Nanotube Growth Conditions via Automated Experimentation. *ACS Nano* **2014**, *8* (10), 10214–10222. https://doi.org/10.1021/nn503347a.

(14) Chang, J.; Nikolaev, P.; Carpena-Núñez, J.; Rao, R.; Decker, K.; Islam, A. E.; Kim, J.; Pitt, M. A.; Myung, J. I.; Maruyama, B. Efficient Closed-Loop Maximization of Carbon Nanotube Growth Rate Using Bayesian Optimization. *Sci Rep* **2020**, *10* (1), 9040. https://doi.org/10.1038/s41598-020-64397-3.

(15) Nikolaev, P.; Hooper, D.; Webber, F.; Rao, R.; Decker, K.; Krein, M.; Poleski, J.; Barto, R.; Maruyama, B. Autonomy in Materials Research: A Case Study in Carbon Nanotube Growth. *NPJ Comput Mater* **2016**, *2* (1), 16031. https://doi.org/10.1038/npjcompumats.2016.31.

(16) Shimizu, R.; Kobayashi, S.; Watanabe, Y.; Ando, Y.; Hitosugi, T. Autonomous Materials Synthesis by Machine Learning and Robotics. *APL Mater* **2020**, *8* (11), 111110. https://doi.org/10.1063/5.0020370.

(17) Kim, C. K.; Drozdov, I. K.; Fujita, K.; Davis, J. C. S.; Božović, I.; Valla, T. In-Situ Angle-Resolved Photoemission Spectroscopy of Copper-Oxide Thin Films Synthesized by Molecular Beam Epitaxy. *J Electron Spectros Relat Phenomena* **2022**, *257*, 146775. https://doi.org/10.1016/j.elspec.2018.07.003.

(18) Furuya, D.; Miyashita, T.; Miura, Y.; Iwasaki, Y.; Kotsugi, M. Autonomous Synthesis System Integrating Theoretical, Informatics, and Experimental Approaches for Large-Magnetic-Anisotropy Materials. *Science and Technology of Advanced Materials: Methods* **2022**, *2* (1), 280–293. https://doi.org/10.1080/27660400.2022.2094698.

(19) Bicer, T.; Gursoy, D.; Kettimuthu, R.; Foster, I. T.; Ren, B.; Andrede, V. De; Carlo, F. De. Real-Time Data Analysis and Autonomous Steering of Synchrotron Light Source Experiments. In *2017 IEEE 13th International Conference on e-Science (e-Science)*; 2017; pp 59–68. https://doi.org/10.1109/eScience.2017.53.

(20) Al-Najjar, A.; Rao, N. S. V; Sankaran, R.; Ziatdinov, M.; Mukherjee, D.; Ovchinnikova, O.; Roccapriore, K.; Lupini, A. R.; Kalinin, S. V. Enabling Autonomous Electron Microscopy for Networked Computation and Steering. In *2022 IEEE 18th International*

*Conference on e-Science (e-Science)*; 2022; pp 267–277. https://doi.org/10.1109/eScience55777.2022.00040.

(21) Taillon, J. A.; Bina, T. F.; Plante, R. L.; Newrock, M. W.; Greene, G. R.; Lau, J. W. NexusLIMS: A Laboratory Information Management System for Shared-Use Electron Microscopy Facilities. *Microscopy and Microanalysis* **2021**, *27* (3), 511–527. https://doi.org/10.1017/S1431927621000222.

(22) Lin, Y. C.; Liu, C.; Yu, Y.; Zarkadoula, E.; Yoon, M.; Puretzky, A. A.; Liang, L.; Kong, X.; Gu, Y.; Strasser, A.; Meyer, H. M.; Lorenz, M.; Chisholm, M. F.; Ivanov, I. N.; Rouleau, C. M.; Duscher, G.; Xiao, K.; Geohegan, D. B. Low Energy Implantation into Transition-Metal Dichalcogenide Monolayers to Form Janus Structures. *ACS Nano* **2020**, *14* (4). https://doi.org/10.1021/acsnano.9b10196.

(23) Harris, S. B.; Lin, Y.-C.; Puretzky, A. A.; Liang, L.; Dyck, O.; Berlijn, T.; Eres, G.; Rouleau, C. M.; Xiao, K.; Geohegan, D. B. Real-Time Diagnostics of 2D Crystal Transformations by Pulsed Laser Deposition: Controlled Synthesis of Janus WSSe Monolayers and Alloys. *ACS Nano* **2023**, *17* (3), 2472–2486. https://doi.org/10.1021/acsnano.2c09952.

(24) Mahjouri-Samani, M.; Tian, M.; Puretzky, A. A.; Chi, M.; Wang, K.; Duscher, G.; Rouleau, C. M.; Eres, G.; Yoon, M.; Lasseter, J.; Xiao, K.; Geohegan, D. B. Nonequilibrium Synthesis of $TiO_2$ Nanoparticle "Building Blocks" for Crystal Growth by Sequential Attachment in Pulsed Laser Deposition. *Nano Lett* **2017**, *17* (8), 4624–4633. https://doi.org/10.1021/acs.nanolett.7b01047.

(25) Shahriari, B.; Swersky, K.; Wang, Z.; Adams, R. P.; Freitas, N. de. Taking the Human Out of the Loop: A Review of Bayesian Optimization. *Proceedings of the IEEE* **2016**, *104* (1), 148–175. https://doi.org/10.1109/JPROC.2015.2494218.

(26) Puretzky, A. A.; Lin, Y.-C.; Liu, C.; Strasser, A. M.; Yu, Y.; Canulescu, S.; Rouleau, C. M.; Xiao, K.; Duscher, G.; Geohegan, D. B. In Situ Laser Reflectivity to Monitor and Control the Nucleation and Growth of Atomically Thin 2D Materials*. *2D Mater* **2020**, *7* (2), 025048. https://doi.org/10.1088/2053-1583/ab7a72.

(27) Kim, H. J.; Chong, M.; Rhee, T. G.; Khim, Y. G.; Jung, M.-H.; Kim, Y.-M.; Jeong, H. Y.; Choi, B. K.; Chang, Y. J. Machine-Learning-Assisted Analysis of Transition Metal Dichalcogenide Thin-Film Growth. *Nano Converg* **2023**, *10* (1), 10. https://doi.org/10.1186/s40580-023-00359-5.

(28) Biswas, A.; Hoyle, C. An Approach to Bayesian Optimization for Design Feasibility Check on Discontinuous Black-Box Functions. *Journal of Mechanical Design* **2021**, *143* (3). https://doi.org/10.1115/1.4049742.

(29) Chu, W.; Ghahramani, Z. Extensions of Gaussian Processes for Ranking: Semisupervised and Active Learning. *NIPS Workshop on Large Scale Kernel Machines*, Whistler 2005. (available at www.merlot.org/merlot/viewMaterial.htm?id=975278)

(30) Held, L.; Holmes, C. C. Bayesian Auxiliary Variable Models for Binary and Multinomial Regression. *Bayesian Anal* **2006**, *1* (1), 145–168. https://doi.org/10.1214/06-BA105.

(31) Wang, Z.; Hutter, F.; Zoghi, M.; Matheson, D.; De Freitas, N. Bayesian Optimization in a Billion Dimensions via Random Embeddings. *J. Artif. Int. Res.* **2016**, *55* (1), 361–387.

(32) Biswas, A.; Vasudevan, R.; Ziatdinov, M.; Kalinin, S. V. Optimizing Training Trajectories in Variational Autoencoders via Latent Bayesian Optimization Approach*. *Mach Learn Sci Technol* **2023**, *4* (1), 015011. https://doi.org/10.1088/2632-2153/acb316.

(33) Biswas, A.; Morozovska, A. N.; Ziatdinov, M.; Eliseev, E. A.; Kalinin, S. V. Multi-Objective Bayesian Optimization of Ferroelectric Materials with Interfacial Control for Memory and Energy Storage Applications. *J Appl Phys* **2021**, *130* (20), 204102. https://doi.org/10.1063/5.0068903

(34) Ueno, T.; Rhone, T. D.; Hou, Z.; Mizoguchi, T.; Tsuda, K. COMBO: An Efficient Bayesian Optimization Library for Materials Science. *Materials Discovery* **2016**, *4*, 18–21. https://doi.org/10.1016/j.md.2016.04.001.

(35) Kalinin, S. V; Ziatdinov, M.; Vasudevan, R. K. Guided Search for Desired Functional Responses via Bayesian Optimization of Generative Model: Hysteresis Loop Shape Engineering in Ferroelectrics. *J Appl Phys* **2020**, *128* (2), 24102. https://doi.org/10.1063/5.0011917

(36) Griffiths, R.-R.; Hernández-Lobato, J. M. Constrained Bayesian Optimization for Automatic Chemical Design. *Chem Sci* **2020**, *11*, 577-586. https://doi.org/10.1039/C9SC04026A

(37) Biswas, A.; Liu, Y.; Creange, N.; Liu, Y.-C.; Jesse, S.; Yang, J.-C.; Kalinin, S. V; Ziatdinov, M. A.; Vasudevan, R. K. A Dynamic Bayesian Optimized Active Recommender System for Curiosity-Driven Human-in-the-Loop Automated Experiments. *arXiv*, April 5, 2023. https://doi.org/10.48550/arXiv.2304.02484 (accessed August 25, 2023).

(38) Frean, M.; Boyle, P. Using Gaussian Processes to Optimize Expensive Functions. In *AI 2008: Advances in Artificial Intelligence*; Wobcke, W., Zhang, M., Eds.; Springer Berlin Heidelberg: Berlin, Heidelberg, 2008; pp 258–267.

(39) Jones, D. R. A Taxonomy of Global Optimization Methods Based on Response Surfaces. *Journal of Global Optimization* **2001**, *21* (4), 345–383. https://doi.org/10.1023/A:1012771025575.

(40) Kushner, H. J. A New Method of Locating the Maximum Point of an Arbitrary Multipeak Curve in the Presence of Noise. *Journal of Basic Engineering* **1964**, *86* (1), 97–106. https://doi.org/10.1115/1.3653121.

(41) Brochu, E.; Cora, V. M.; de Freitas, N. A Tutorial on Bayesian Optimization of Expensive Cost Functions, with Application to Active User Modeling and Hierarchical Reinforcement Learning. *arXiv*, December 12, 2010. https://doi.org/10.48550/arXiv.1012.2599 (accessed August 25, 2023).

(42) Cox, D. D.; John, S. A Statistical Method for Global Optimization. In *Proceedings of the 1992 IEEE International Conference on Systems, Man, and Cybernetics*; Chicago, IL, USA, 18–21 October 1992; pp.1241–1246. https://doi.org/10.1109/ICSMC.1992.271617.

(43) Zhang, Y.; Guo, H.; Sun, W.; Sun, H.; Ali, S.; Zhang, Z.; Saito, R.; Yang, T. Scaling Law for Strain Dependence of Raman Spectra in Transition-Metal Dichalcogenides. *Journal of Raman Spectroscopy* **2020**, *51* (8), 1353–1361. https://doi.org/10.1002/jrs.5908.

(44) Zhang, X.; Qiao, X.-F.; Shi, W.; Wu, J.-B.; Jiang, D.-S.; Tan, P.-H. Phonon and Raman Scattering of Two-Dimensional Transition Metal Dichalcogenides from Monolayer,

Multilayer to Bulk Material. *Chem Soc Rev* **2015**, *44* (9), 2757–2785. https://doi.org/10.1039/C4CS00282B.

(45) Mahjouri-Samani, M.; Liang, L.; Oyedele, A.; Kim, Y.-S.; Tian, M.; Cross, N.; Wang, K.; Lin, M.-W.; Boulesbaa, A.; Rouleau, C. M.; Puretzky, A. A.; Xiao, K.; Yoon, M.; Eres, G.; Duscher, G.; Sumpter, B. G.; Geohegan, D. B. Tailoring Vacancies Far Beyond Intrinsic Levels Changes the Carrier Type and Optical Response in Monolayer $MoSe_{2-x}$ Crystals. *Nano Lett* **2016**, *16* (8), 5213–5220. https://doi.org/10.1021/acs.nanolett.6b02263.

(46) Stanford, M. G.; Pudasaini, P. R.; Belianinov, A.; Cross, N.; Noh, J. H.; Koehler, M. R.; Mandrus, D. G.; Duscher, G.; Rondinone, A. J.; Ivanov, I. N.; Ward, T. Z.; Rack, P. D. Focused Helium-Ion Beam Irradiation Effects on Electrical Transport Properties of Few-Layer $WSe_2$: Enabling Nanoscale Direct Write Homo-Junctions. *Sci Rep* **2016**, *6* (1), 27276. https://doi.org/10.1038/srep27276.

(47) Fang, L.; Chen, H.; Yuan, X.; Huang, H.; Chen, G.; Li, L.; Ding, J.; He, J.; Tao, S. Quick Optical Identification of the Defect Formation in Monolayer $WSe_2$ for Growth Optimization. *Nanoscale Res Lett* **2019**, *14* (1), 274. https://doi.org/10.1186/s11671-019-3110-z.

(48) Virtanen, P.; Gommers, R.; Oliphant, T. E.; Haberland, M.; Reddy, T.; Cournapeau, D.; Burovski, E.; Peterson, P.; Weckesser, W.; Bright, J.; van der Walt, S. J.; Brett, M.; Wilson, J.; Millman, K. J.; Mayorov, N.; Nelson, A. R. J.; Jones, E.; Kern, R.; Larson, E.; Carey, C. J.; Polat, \.Ilhan; Feng, Y.; Moore, E. W.; VanderPlas, J.; Laxalde, D.; Perktold, J.; Cimrman, R.; Henriksen, I.; Quintero, E. A.; Harris, C. R.; Archibald, A. M.; Ribeiro,

A. H.; Pedregosa, F.; van Mulbregt, P.; SciPy 1.0 Contributors. SciPy 1.0: Fundamental Algorithms for Scientific in Python. *Nat Methods* **2020**, *17*, 261–272. https://doi.org/10.1038/s41592-019-0686-2.

(49) Rodríguez-Martínez, X.; Vezie, M. S.; Shi, X.; McCulloch, I.; Nelson, J.; Goñi, A. R.; Campoy-Quiles, M. Quantifying Local Thickness and Composition in Thin Films of Organic Photovoltaic Blends by Raman Scattering. *J Mater Chem C Mater* **2017**, *5* (29), 7270–7282. https://doi.org/10.1039/C7TC01472D.

(50) Seo, S.; Choi, H.; Kim, S.-Y.; Lee, J.; Kim, K.; Yoon, S.; Lee, B. H.; Lee, S. Growth of Centimeter-Scale Monolayer and Few-Layer $WSe_2$ Thin Films on $SiO_2$/Si Substrate via Pulsed Laser Deposition. *Adv Mater Interfaces* **2018**, *5* (20), 1800524. https://doi.org/10.1002/admi.201800524.

(51) Seo, S.; Kim, S.; Choi, H.; Lee, J.; Yoon, H.; Piao, G.; Park, J.-C.; Jung, Y.; Song, J.; Jeong, S. Y.; Park, H.; Lee, S. Direct In Situ Growth of Centimeter-Scale Multi-Heterojunction $MoS_2$/$WS_2$/$WSe_2$ Thin-Film Catalyst for Photo-Electrochemical Hydrogen Evolution. *Advanced Science* **2019**, *6* (13), 1900301. https://doi.org/10.1002/advs.201900301.

(52) Serna, M. I.; Yoo, S. H.; Moreno, S.; Xi, Y.; Oviedo, J. P.; Choi, H.; Alshareef, H. N.; Kim, M. J.; Minary-Jolandan, M.; Quevedo-Lopez, M. A. Large-Area Deposition of $MoS_2$ by Pulsed Laser Deposition with In Situ Thickness Control. *ACS Nano* **2016**, *10* (6), 6054–6061. https://doi.org/10.1021/acsnano.6b01636.

(53) Serrao, C. R.; Diamond, A. M.; Hsu, S.-L.; You, L.; Gadgil, S.; Clarkson, J.; Carraro, C.; Maboudian, R.; Hu, C.; Salahuddin, S. Highly Crystalline $MoS_2$ Thin Films Grown by

Pulsed Laser Deposition. *Appl Phys Lett* **2015**, *106* (5), 052101. https://doi.org/10.1063/1.4907169.

(54) Lowndes, D. H.; Geohegan, D. B.; Puretzky, A. A.; Norton, D. P.; Rouleau, C. M. Synthesis of Novel Thin-Film Materials by Pulsed Laser Deposition. *Science* **1996**, *273* (5277), 898–903. https://doi.org/10.1126/science.273.5277.898.

(55) Jellison, G. E.; Modine, F. A. Optical Functions of Silicon at Elevated Temperatures. *J Appl Phys* **1994**, *76* (6), 3758–3761. https://doi.org/10.1063/1.357378.

(56) Jung, G.-H.; Yoo, S.; Park, Q.-H. Measuring the Optical Permittivity of Two-Dimensional Materials without a Priori Knowledge of Electronic Transitions. *Nanophotonics* **2019**, *8* (2), 263–270. https://doi.org/doi:10.1515/nanoph-2018-0120.

Supporting Information

for

# Autonomous synthesis of thin film materials with pulsed laser deposition enabled by in situ spectroscopy and automation

*Sumner B. Harris\*, Arpan Biswas, Seok Joon Yun, Christopher M. Rouleau, Alexander A. Puretzky, Rama K. Vasudevan, David B. Geohegan, Kai Xiao\**

Center for Nanophase Materials Sciences, Oak Ridge National Laboratory, Oak Ridge, Tennessee 37831, United States.

\*Correspondence should be addressed to: harrissb@ornl.gov or xiaok@ornl.gov

Notice: This manuscript has been authored by UT-Battelle, LLC, under Contract No. DE-AC05- 00OR22725 with the U.S. Department of Energy. The United States Government retains and the publisher, by accepting the article for publication, acknowledges that the United States Government retains a non-exclusive, paid-up, irrevocable, world-wide license to publish or reproduce the published form of this manuscript, or allow others to do so, for United States Government purposes. The Department of Energy will provide public access to these results of federally sponsored research in accordance with the DOE Public Access Plan ( http://energy.gov/downloads/doe-public-access-plan ).

# Table of contents

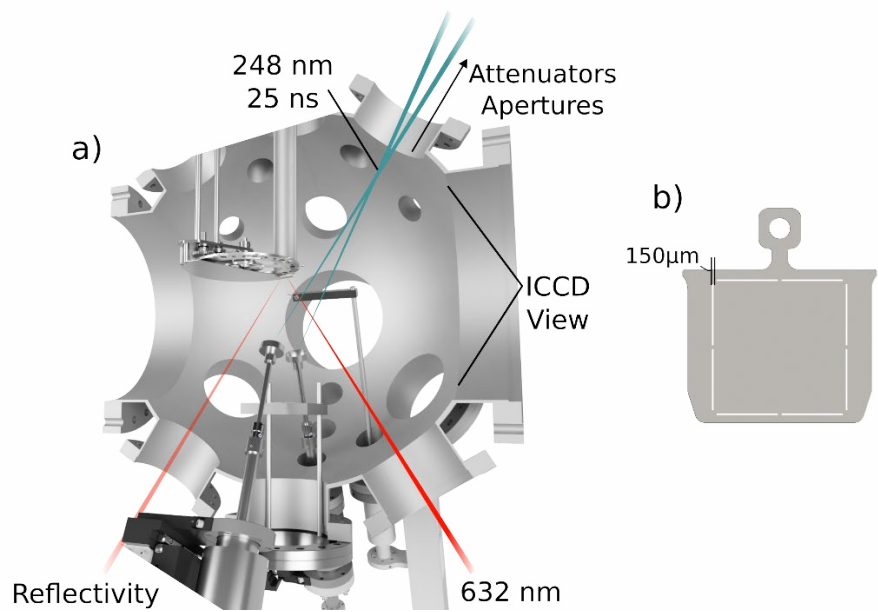

**Figure S1. Additional details of the Auto-PLD system.** a) The top view cross section of the chamber (X-Z plane) shows the 248nm KrF excimer beam paths and the 632nm HeNe beam path used for real-time laser reflectivity. b) The sample plate is the standard flag-style design but include a custom laser-cut, 150 μm wide, thermal break to minimize the heat flow from the sample plates to the substrate wheel.

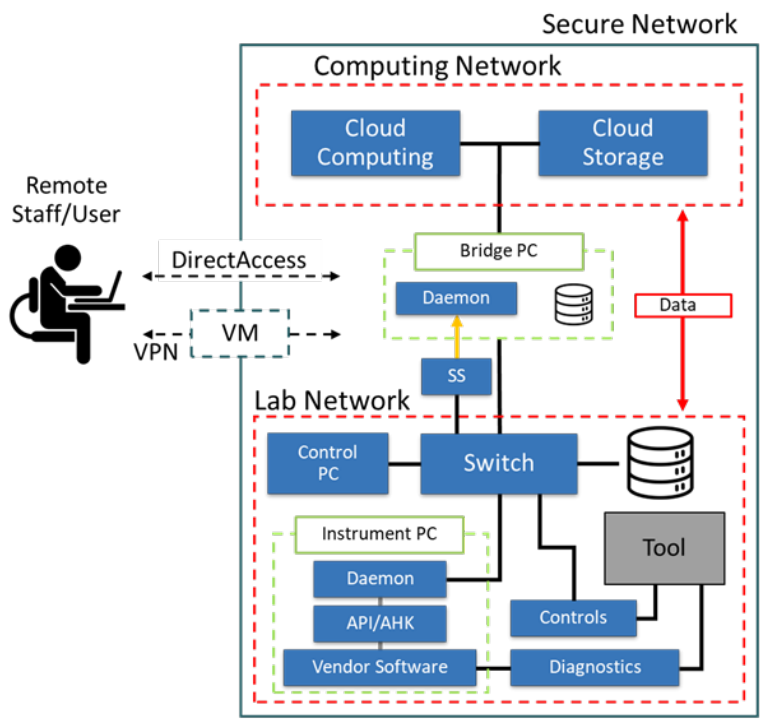

**Figure S2. The laboratory network ecosystem that links the Auto-PLD to cloud computing, storage resources, and enables remote access.** Within an institutionally maintained secure network, independent lab networks are linked through a bridge PC for data transfer and remote access. Synthesis and characterization instruments, in this case the Auto-PLD (Tool), are configured for ethernet control using local daemon server programs using an API or AHK when direct hardware control is not possible. Messages/commands from the lab network are sent to the bridge PC via serial server (SS) to transfer data or execute cloud programs.

**Note S1. Description of laboratory network eco system**

Auto-PLD is integrated into a larger laboratory ecosystem (we call it *Sci-by-Wire*) which provides a direct link between multiple synthesis/characterization labs, cloud resources such as HPC or storage, and allows remote access to the equipment. Figure S2 shows a schematic of the network infrastructure used to enable this connectivity. Within our secure, institutionally maintained network, we set up a virtual local area network (VLAN) to maintain a persistent software/operating system state on instrument computers while still allowing automated ingress and egress of data between the VLAN and the cloud networks. All instruments and tools within the lab network are interfaced through ethernet and can be independently controlled by TCP/IP using any machine with secure network access. Users outside of the secure network can be granted remote access through a virtual machine and institutional authentication.

Autonomous experiments are orchestrated with open-source Python software which allows for seamless integration of popular AI/ML libraries. Direct communication is possible with most instruments in the ecosystem but some must be interacted with through vendor software on independent PCs. In these instances, we run daemon server programs on the instrument PCs which can pass messages and commands between the vendor PC and the experiment control program. The daemons command instruments through an applied programming interface (API) or, when absolutely required, automated keystrokes and mouse clicks, e.g. AutoHotkey (AHK). Data transfer between the VLAN and cloud network is done through a purpose-built bridge server which uses also uses a daemon to pass commands for file transfer and script execution on the cloud. To effectively manage the multimodal datasets that are generated, all the process parameters, plume diagnostics, and characterization data captured for each sample are compiled into a single Hierarchical Data Format (HDF5) file and uploaded to local/cloud storage to create an easily searchable database.

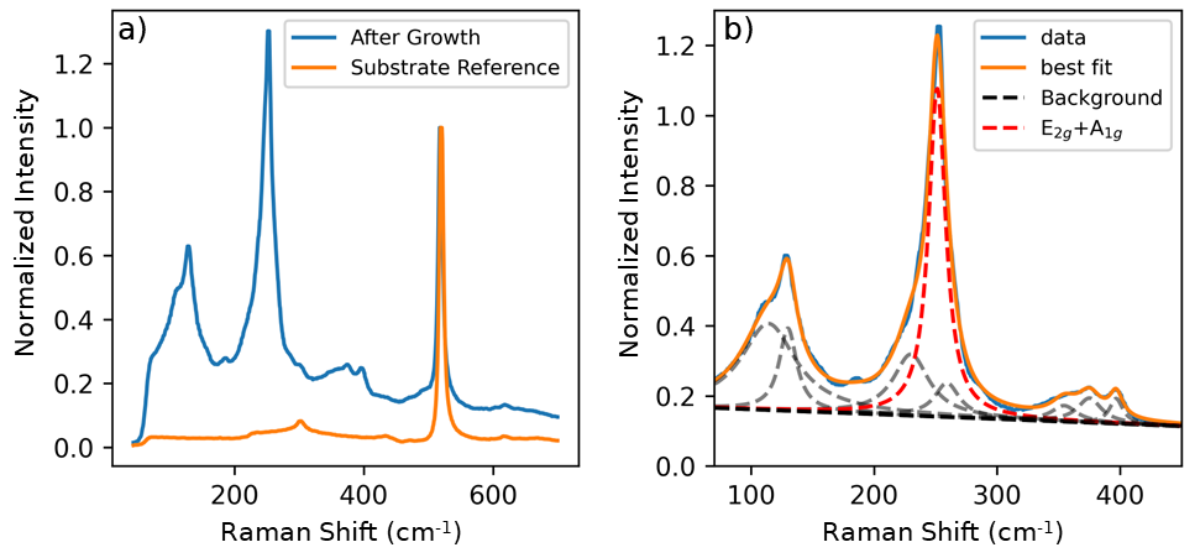

**Figure S3. Example of fitting procedure to determine the Raman score.** a) The Raman spectra of the substrate before growth and the sample after-growth are both normalized to the Si peak at 520.7 cm$^{-1}$. b) The reference spectrum is subtracted from the data and the same multi-Lorentzian fitting routine is used to extract the prominence and FWHM of the WSe$_2$ E$_{2g}$+A$_{1g}$ peak for every sample.

**Note S2. Detailed description of the autonomous workflow used in the present study.**

The detailed workflow of the autonomous PLD experiment is as follows. First, a set of 4 parameters (P, T, $F_1$, $F_2$) are chosen by the BO and passed to the Auto-PLD. The substrate wheel rotates a fresh substrate into place, the sample file is initialized with the sample name and metadata related to the fixed growth parameters, the background pressure is set and stabilized, a room temperature Raman spectrum is collected for reference, and then the substrate is heated. Once the substrate temperature has stabilized, a high temperature reference Raman spectrum is collected, laser reflectivity acquisition is started, and the deposition begins. During the deposition, a series of 50 ICCD images and an ion probe trace are collected. Once the target reflectivity contrast value is reached (or a time-out of 45 minutes), the deposition is stopped and a high temperature, post-growth Raman spectrum is collected. The sample temperature is ramped down and then allowed to cool ambiently for 10 minutes to reach near room temperature. After the cooling period, a final Raman spectrum is collected, and the sample score is calculated. The existing dataset is augmented to include the new P, T, $F_1$, $F_2$, with the associated score, the BO is updated and trained, and outputs the next suggested set of growth parameters. This process repeats a maximum of 10 times, the number of substrates on the wheel, without human intervention. Once 10 samples have been grown, the chamber is manually vented, new substrates are mounted on the wheel, the laser window is cleaned, and the chamber is pumped back down to continue the autonomous routine.

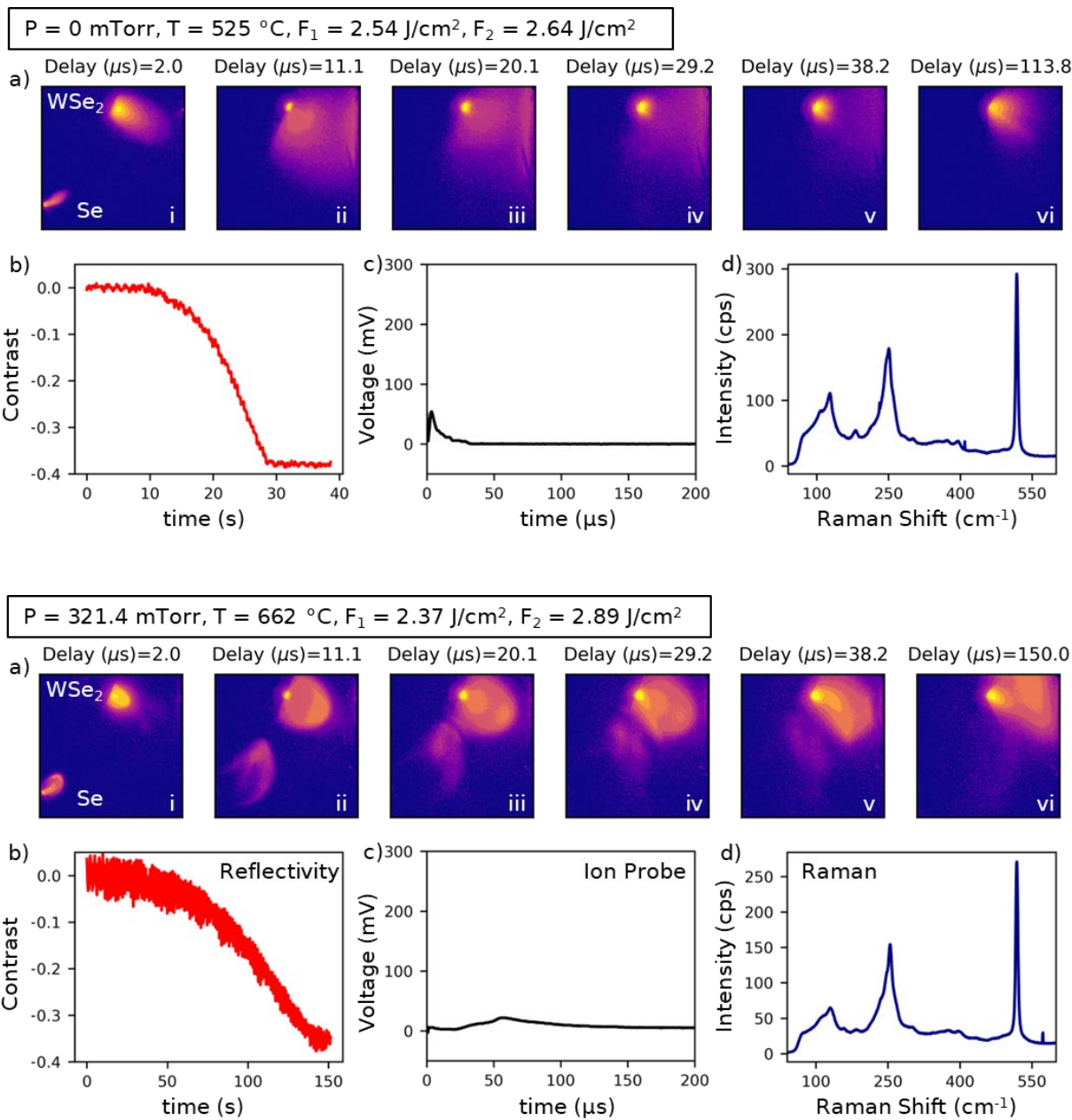

**Figure S4.1. Examples of *in situ* diagnostics collected for two samples (top and bottom) grown with high laser fluences at different pressures with similar outcomes.** a) ICCD images taken at various delay times with the gate set for 10% of the delay. b) Laser reflectivity contrast vs time. c) Ion probe signal. d) Room temperature Raman spectrum collected after growth, from which the sample score was calculated. These 2 samples were grown with similar, high fluence conditions but a vacuum (top) and 321 mTorr (bottom). The ICCD images (a) show plume mixing at vacuum but more hydrodynamic like interactions at high pressure with the $WSe_2$ being plume deflected by the weakly luminous Se plume and longer delay times (150 μs). The growth time is 6.5x slower at high pressure (130s vs 20s) and the ion probe indicates significant slowing of the plume.

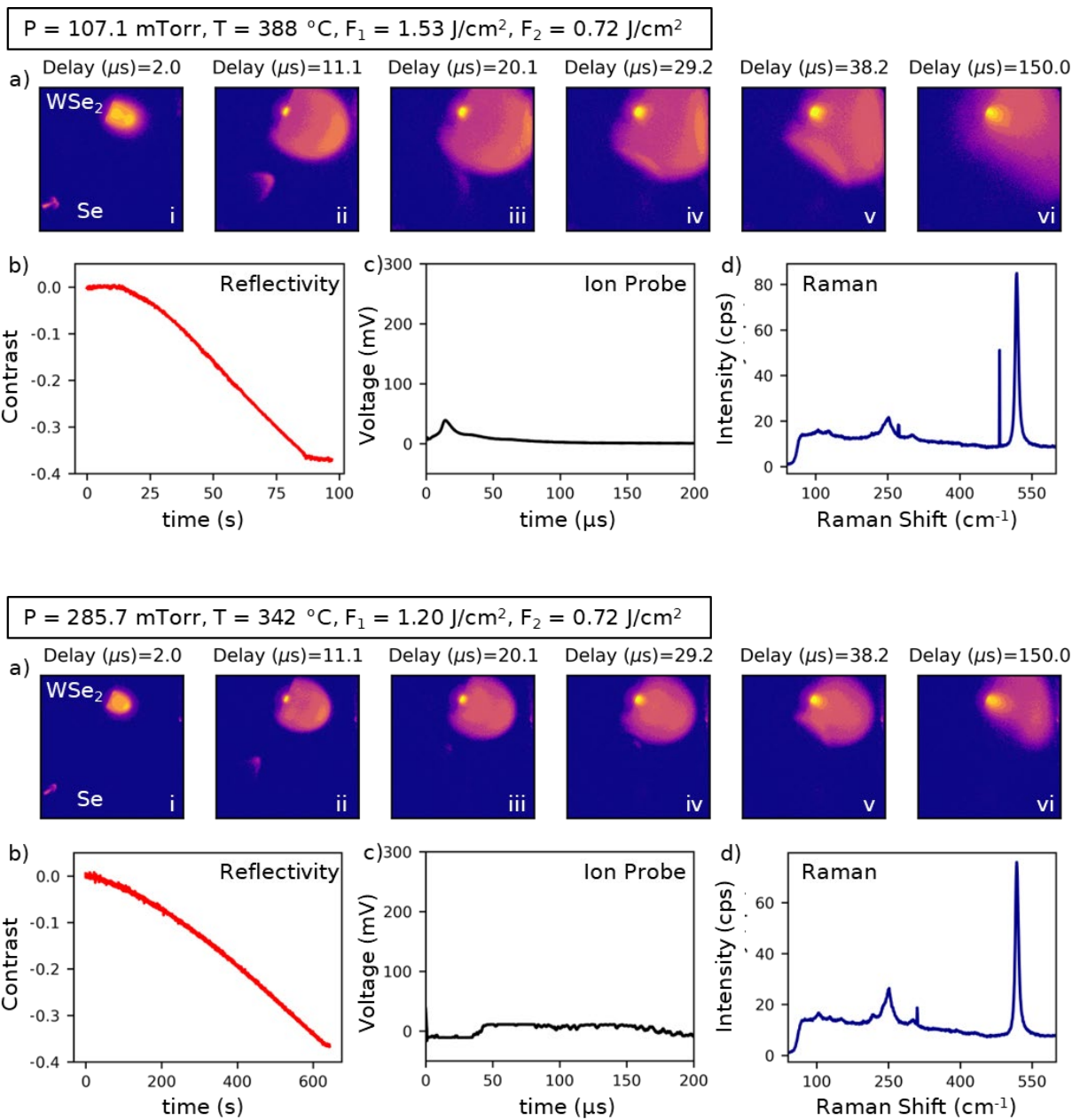

**Figure S4.2. Examples of *in situ* diagnostics collected for two samples (top and bottom) grown with low laser fluence and low temperature at moderate pressures.** a) ICCD images taken at various delay times with the gate set for 10% of the delay. b) Laser reflectivity contrast vs time. c) Ion probe signal. d) Room temperature Raman spectrum collected after growth, from which the sample score was calculated. These 2 samples were grown with similar low laser fluences and temperatures, both have poor outcomes based on the Raman spectra. Based on the ICCD images, the higher pressure (bottom) leads to more WSe$_2$ plume confinement and slowing, with the leading edge reaching the substrate at ~ 50 μs vs the lower pressure (top) where the plume reaches the substrate at ~ 15 μs.

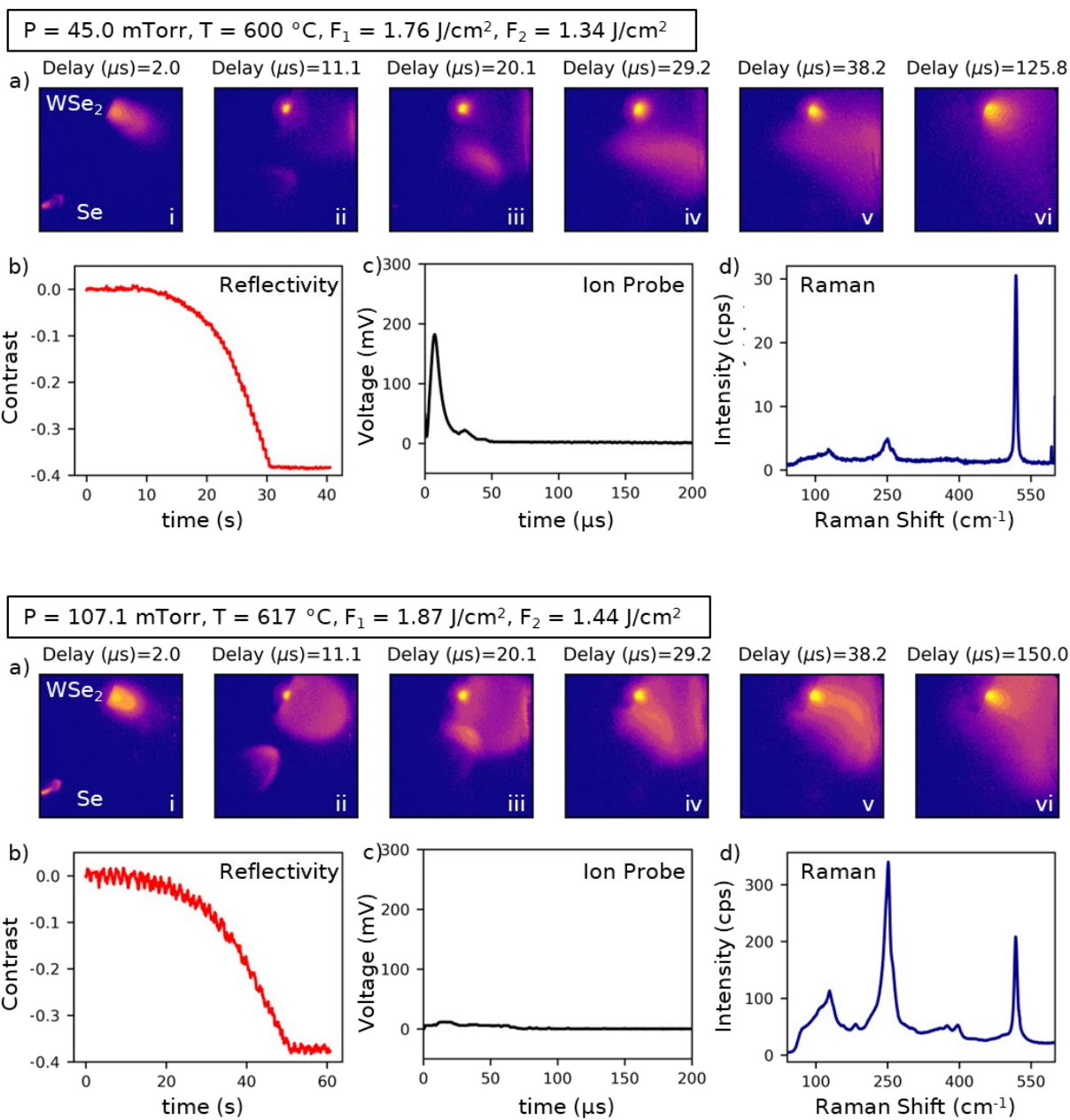

**Figure S4.3. Examples of in situ diagnostics collected for two samples (top and bottom) grown with similar fluence and temperature at different pressures with different outcomes.** a) ICCD images taken at various delay times with the gate set for 10% of the delay. b) Laser reflectivity contrast vs time. c) Ion probe signal. d) Room temperature Raman spectrum collected after growth, from which the sample score was calculated. These two samples differ mainly in the background pressure. The sample grown at 45 mTorr (top) gives a poor Raman score but increasing the pressure to 107 mTorr (top) significantly improves the result. The growth times are similar (20 s for 45 mTorr and 30s for 107 mTorr) but the plume dynamics are different. The ion probe shows a high current of fast species at 45mTorr that are attenuated and slowed at 107 mTorr.

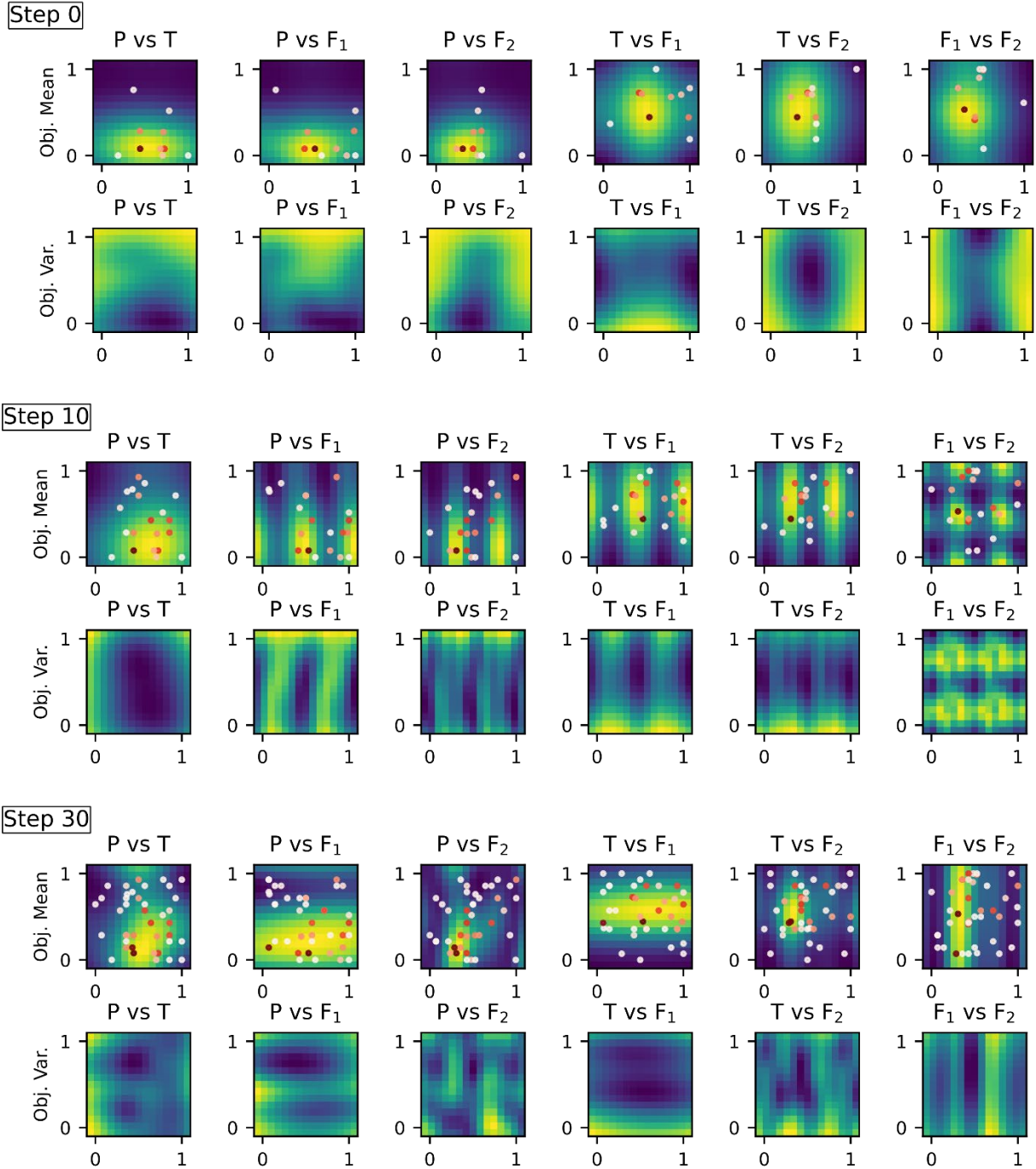

**Figure S5.1. BO evolution for steps 0 (top), 10 (middle) and 30 (bottom).** The averaged surrogate function (Obj. Mean) and the averaged variance (Obj. Var.) are projected into each 2D parameter plane for visualization. The parameter axes are displayed in normalized units which was used during the active learning steps. As the BO is updated with more samples, a periodic structure can be seen in the surrogate function because of the periodic kernel.

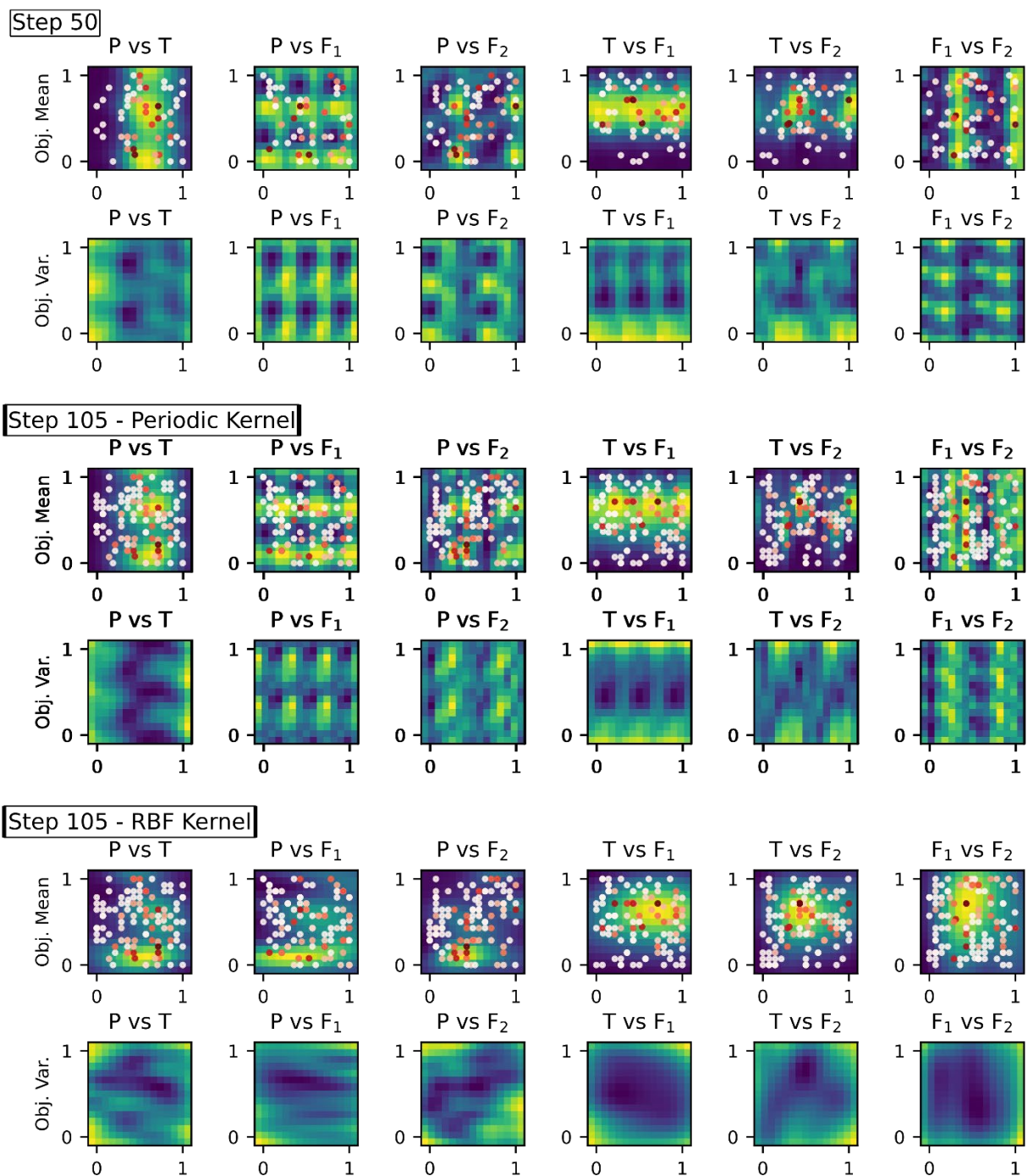

**Figure S5.2. BO evolution for steps 50 (top), 105 as calculated with the periodic kernel (middle) and 105 calculated with the RBF kernel (bottom).** The averaged surrogate function (Obj. Mean) and the averaged variance (Obj. Var.) are projected into each 2D parameter plane for visualization. The parameter axes are displayed in normalized units which was used during the active learning steps. The BO evolved very little between steps 50 and 105 with the periodic kernel. Switching to the RBF kernel after step 105 yielded a more interpretable, physical surrogate without biasing the parameter space exploration to the edges of the domain.

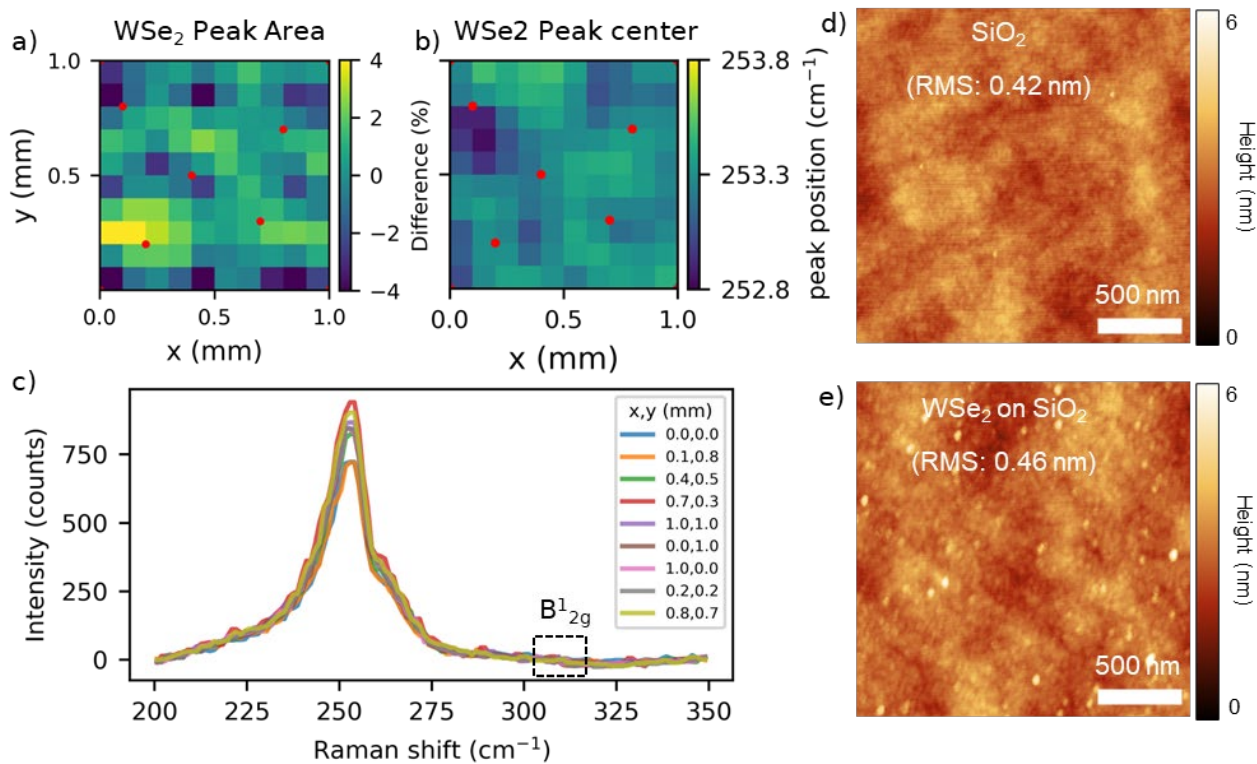

**Figure S6. Raman map over a 1x1 mm² area and AFM scan of a typical, successful WSe₂ film**. a) The WSe$_2$ E$_{2g}$+A$_{1g}$ peak area's difference from the average is ± 4% and the b) peak center has a standard deviation of ±0.3 cm$^{-1}$, indicating that the film is uniform over a 1 mm² area. c) Individual spectra of the WSe$_2$ E$_{2g}$+A$_{1g}$ taken from the red points shown in a-b). Notably, the B$^1_{2g}$ peak at 310 cm$^{-1}$ is absent at every point, indicating that the film does not have any significant multilayer regions. AFM scans show an RMS surface roughness of 0.46 nm which is comparable to the Si substrate's roughness of 0.42 nm.

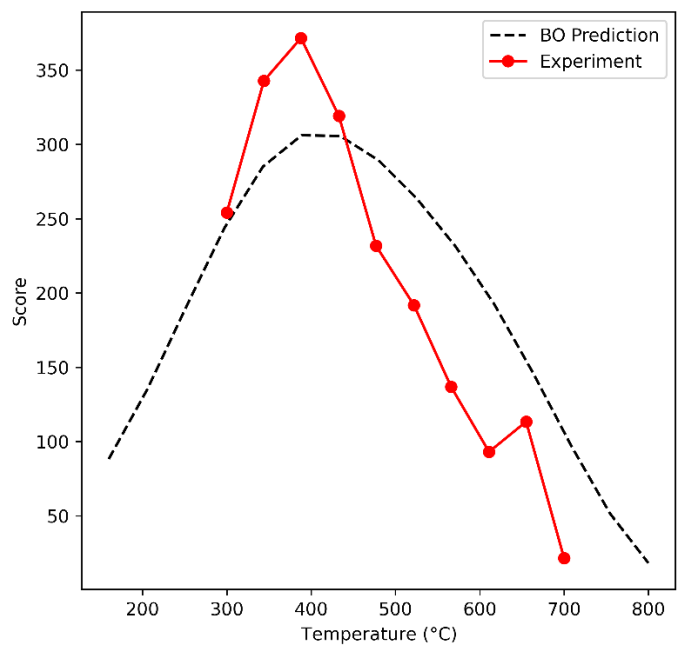

**Figure S7. BO prediction of score vs temperature along an axis that has only 1 nearby sample from the full 125 sample experiment.** After the 125 active learning steps, a series of 10 samples were grown where only the temperature was varied and the other parameters were fixed to P = 86.2 mTorr, $F_1$ = 1.08 J/cm$^2$, and $F_2$ = 1.39 J/cm$^2$ without updating the BO. The BO prediction for Raman score along this unexplored axis in parameter space qualitatively matches the experimental outcome.